\documentclass{aa}

\usepackage{color}

\newcommand{\D}{{\rm d}}

\newcommand{\rme}{{\rm e}}

\newcommand{\nii}{[\ion{N}{ii}]}
\newcommand{\oiii}{[\ion{O}{iii}]}
\newcommand{\Ha}{H$\alpha$}
\newcommand{\HaLF}{H$\alpha$LF}

\usepackage{graphicx}
\usepackage{txfonts}
\begin{document} 

   \title{Modelling the number density of H$\alpha$ emitters\\ for future spectroscopic near-IR space missions}

   \author{
	Pozzetti, L. \inst{1},
	Hirata, C.M. \inst{2},
	Geach, J.E. \inst{3},
	Cimatti, A. \inst{4}, 
	Baugh, C. \inst{5},
	Cucciati, O. \inst{1,4},
	Merson A. \inst{6},
	Norberg P. \inst{5},
	Shi D. \inst{5},
          }

   \institute{ 
INAF Osservatorio Astronomico di Bologna, Via Ranzani 1, I-40127 Bologna, Italy
\and 
Center for Cosmology and Astroparticle Physics, The Ohio State University, 191 West Woodruff Lane, Columbus, Ohio 43210, USA
\and 
Centre for Astrophysics Research, Science and Technology Research Institute, University of Hertfordshire, Hatfield AL10 9AB, UK
\and 
Dipartimento di Fisica e Astronomia, Universit\`a di Bologna, viale Berti Pichat 6/2, 40127 Bologna, Italy
\and 
Institute for Computational Cosmology (ICC), Department of Physics, Durham University, South Road, Durham, DH1 3LE
\and 
Department of Physics and Astronomy, University College London, Gower Street, London, WC1E 6BT
             }

\titlerunning{Modelling the number density of H$\alpha$ emitters}
\authorrunning{Pozzetti et al.}

   \date{4 March 2016}

 
  \abstract
   {The future space missions {\slshape Euclid} and {\slshape WFIRST-AFTA}  
will use the H$\alpha$ emission line to measure the redshifts of tens of millions of galaxies. The H$\alpha$ luminosity function at $z>0.7$ is one of the major sources of uncertainty in forecasting cosmological constraints from these missions.}
   {We construct unified empirical models of the H$\alpha$ luminosity function spanning the range of redshifts and line luminosities relevant to the redshift surveys proposed with {\slshape Euclid} and {\slshape WFIRST-AFTA}.}
   {By fitting to observed luminosity functions from H$\alpha$ surveys, we build three models for its evolution. Different fitting methodologies, functional forms for the luminosity function, subsets of the empirical input data, and treatment of systematic errors are considered to explore the robustness of the results.}
   {Functional forms and model parameters are provided for all three models, along with the counts and redshift distributions up to $z\sim2.5$ for a range of limiting fluxes ($F_{{\rm H}\alpha}>0.5 - 3\times 10^{-16}$ erg cm$^{-2}$ s$^{-1}$) that are relevant for future space missions. 
For instance, in the redshift range $0.90<z<1.8$, our models predict an available galaxy density in the range 7700--13300 and 2000--4800 deg$^{-2}$ respectively at fluxes above $F_{{\rm H}\alpha}>1$ and $2\times 10^{-16}$ erg cm$^{-2}$ s$^{-1}$, and 32000--48000 for $F_{{\rm H}\alpha}>0.5 \times 10^{-16}$ erg cm$^{-2}$ s$^{-1}$ in the extended redshift range $0.40<z<1.8$.
We also consider the implications of our empirical models for the total \Ha\ luminosity density of the Universe, and the closely related cosmic star formation history.}
   {}


   \maketitle
%

\section{Introduction}

Since the discovery of the apparent acceleration of the expansion of the Universe \citep[e.g.][]{1998AJ....116.1009R, 1999ApJ...517..565P}, many efforts have been made to measure the dark energy equation of state,
exploiting different observations. Among the suggestions proposed, the use of baryon acoustic oscillations (BAO) as standard rulers appears to have a particularly low level of systematic uncertainty since it corresponds to a feature in the correlation function, whereas most observational and astrophysical systematics are expected to be broad-band \citep[e.g.][]{2006astro.ph..9591A}. Indeed, in recent years, the BAO technique has seen a dramatic improvement in capability owing to the increase in volume probed by galaxy surveys \citep[e.g.][]{2005MNRAS.362..505C, 2005ApJ...633..560E, 2007MNRAS.381.1053P, 2010MNRAS.401.2148P, 2011MNRAS.415.2892B, 2011MNRAS.418.1707B, 2012MNRAS.427.2132P, 2012MNRAS.427.2146X, 2013MNRAS.435...64K, 2014MNRAS.441...24A, 2014MNRAS.441.3524K}.

The future space-based galaxy redshift surveys planned for the ESA's {\slshape Euclid} \citep{2011arXiv1110.3193L} and NASA's Wide-Field Infrared Survey Telescope - Astrophysics Focused Telescope Assets design ({\slshape WFIRST-AFTA}) \citep[]{Spergel15, Green12} missions will use near-IR (NIR) slitless spectroscopy to collect large samples of emission-line galaxies to probe dark energy. These spectroscopic surveys will identify mainly \Ha\ emitters out as far as $z\sim 2$, and their maps of large-scale structure will be used for studies of BAO, power spectrum $P(k)$ in general, large-scale structure, as well as other statistics such as the measurement of the rate of growth of structure using redshift space distortions \citep{1987MNRAS.227....1K, 2008Natur.451..541G}. In this context the space density of \Ha\ emitters (i.e. their luminosity function) is a key ingredient for a mission's 
performance forecast to determine the number of objects above the mission's sensitivity threshold and 
optimize the survey.

It is known that the cosmic star formation rate was higher in the Universe's past than it is today,
possibly peaking near $z\sim2$ \citep{2014ARA&A..52..415M}, thereby
ensuring a high number of star-forming objects with high luminosity at high redshift suitable for BAO measurements.
However, the abundance of \Ha\ emitters detectable by blind spectroscopy has historically been firmly established only at low redshift by means of spectroscopic surveys in the optical
\citep[e.g.][]{1995ApJ...455L...1G}.
At higher redshift, from the ground, the intense airglow makes NIR spectroscopic searches for emission line galaxies impractical; thus systematic
ground-based NIR \Ha\ spectroscopic searches in early studies have been limited to small areas with single slit spectroscopy \citep[e.g.][]{2002MNRAS.337..369T}.
Therefore,
narrow-band NIR searches have been used as an alternative method 
for identifying large numbers of $z>0.7$ emission-line galaxies, e.g. HiZELS \citep{2008MNRAS.388.1473G, 2009MNRAS.398...75S, 2012MNRAS.420.1926S, 2013MNRAS.428.1128S} and the NEWFIRM \Ha\ Survey \citep{2011ApJ...726..109L}.
These surveys have the advantage of wide area and high sensitivity to emission lines 
but suffer from their narrow redshift ranges and significant contamination from emission lines at different redshifts. 
From space, grism spectroscopy with NICMOS \citep{1999ApJ...519L..47Y, 2009ApJ...696..785S} and more recently with the Wide Field Camera 3 (WFC3) on the {\slshape Hubble Space Telescope}, have allowed small-area 
surveys, such as the
WFC3 Infrared Spectroscopic Parallels (WISP) survey \citep{2013ApJ...779...34C},
at relevant fluxes (deeper than {\slshape Euclid} or {\slshape WFIRST-AFTA}) to probe the luminosity function of emission line objects at high redshift.

As a result, early studies of space-based galaxy redshift surveys often based their \Ha\ luminosity function models on indirect extrapolations from alternative star formation indicators such as the rest-frame ultraviolet continuum or [\ion{O}{ii}] line strength \citep[e.g.][]{2007ApJ...657..738L, 2007ApJS..172..456T, 2008ApJS..175...48R}. While physically motivated, this procedure suffers from a multitude of uncertainties in the details of the \ion{H}{ii} region parameters, dust extinction, stellar populations, and the joint distribution thereof, and the impact of uncertainties in the predicted \Ha\ flux is enhanced by the steepness of the luminosity function.

Motivated by the prospect of future dark energy surveys targeting \Ha\ emitters at near-infrared wavelengths 
(i.e. $z > 0.5$), \citet{2010MNRAS.402.1330G} used the empirical data available at that time to model the evolution of 
the \Ha\ luminosity function out to $z\sim2$. Much more ground- and space-based data have become available since then, thanks largely to the same improvements in NIR detector technology that make {\slshape Euclid} and {\slshape WFIRST-AFTA} possible. 
In particular, the WFC3 Infrared Spectroscopic Parallels (WISP) survey \citep{2013ApJ...779...34C} has enabled the blind detection of large numbers of \Ha\ emitters. Due to its similarity to the observational setups planned for {\slshape Euclid} and {\slshape WFIRST-AFTA}, it is an excellent test case against which to calibrate expectations for these future missions.

In this work, we update the empirical model of \citet{2010MNRAS.402.1330G}, collecting a larger dataset of \Ha\ luminosity functions from low- to high-redshift, in order to constrain the evolution of the space density of \Ha\ emitters. We construct three empirical models 
and make prediction for future \Ha\ surveys as a function of sensitivity threshold (i.e. counts) and redshift (i.e. redshift distributions).
We scale all the luminosity functions to a reference cosmology with $H_0=70$ km s$^{-1}$ Mpc$^{-1}$ and $\Omega_{\rm m}=0.3$, and present results in terms of comoving volume. 
The models and luminosity functions presented here are for \Ha\ only, \emph{not} \Ha+\nii.
The final aim of these models is to provide key inputs for instrumental simulations
essential to derive forecast in 
future space missions, like {\slshape Euclid} and {\slshape WFIRST-AFTA}, 
that at the nominal resolution will be partially able to resolve \Ha.
Future simulations will clarify all the observational effects, 
from source confusion to the \nii\ contamination
and percentiles of blended lines, completeness and selection effects.

We emphasize here that our models are empirical and therefore
we have reduced as much as possible any astrophysics assumption, but those 
based on \Ha\ public data.
Furthermore, we do not attempt to exclude the AGN contribution from the bright end of the \Ha\ public
luminosity functions, being AGNs valid sources (as \Ha\ emitters) for the current planned missions.
Thus they should not be excluded
to derive the total number of \Ha\ emitters mapped by {\slshape Euclid} and {\slshape WFIRST-AFTA}.

This paper is organized as follows. The input data is described in \S\ref{sec:empirical}. The three models and the procedures used to derive them are described in \S\ref{sec:models}, and the comparison to the input data is summarized in \S\ref{sec:compare-e}. \S\ref{sec:abundance} describes the redshift and flux distribution, with a focus on the ranges relevant to {\slshape Euclid} and {\slshape WFIRST-AFTA}, and \S\ref{sec:mocks} compares our results to 
semi-analytic mock catalogues. Our \Ha\ luminosity functions are compared to other estimates of the cosmic star formation history in \S\ref{sec:sfh}. We conclude in \S\ref{sec:discussion}. Technical details are placed in the appendices.

\section{Empirical luminosity functions} 
\label{sec:empirical}

Forecasts for future NIR slitless galaxy redshift surveys 
require as input the luminosity function of H$\alpha$ emitters (H$\alpha$LF) in order to determine the number of objects above the mission's sensitivity threshold. 
In particular, we focus on the prediction for the originally planned {\slshape Euclid} Wide grism survey \citep{2011arXiv1110.3193L}, i.e  to a flux limit $F_{{\rm H}\alpha}> 3 \times 10^{-16}$ erg cm$^{-2}$ s$^{-1}$, and  detectore sensitivity $1.1<\lambda<2.0\ \mu$m (sampling H$\alpha$ at $0.70<z<2.0$) over 15,000 deg$^2$. A smaller area (2200 deg$^2$) and a fainter flux limit ($> 1 \times 10^{-16}$ erg cm$^{-2}$ s$^{-1}$) is the baseline depth of {\slshape WFIRST-AFTA} \citep{Green11} with a grism spanning the wavelength range $1.35-1.95$ $\mu$m.

The original predictions presented in the {\slshape Euclid} Definition Study Report \citep[the Red Book,][]{2011arXiv1110.3193L} used the predicted counts of \Ha\ emitters by \citet{2010MNRAS.402.1330G}. 
This model was based on {\slshape Hubble Space Telescope} and other data available prior to 2010. 
Here, we provide an updated compilation of empirical H$\alpha$ LFs available in literature, and use 
the most recent and verified ones out to $z_{\rm max}\sim 2.3$ to build three updated models of \Ha\ emitters counts.
To provide precise predictions over the redshift range of interest of NIR missions (i.e. $0.7<z<2$) all three models include estimates from the HiZELS narrow-band ground-based imaging survey with UKIRT, Subaru and VLT \citep{2013MNRAS.428.1128S}, covering $\sim$ 2 deg$^2$ in the COSMOS field at $z=0.4,0.84,1.47,2.23$; the WISP slitless space-based spectroscopic survey with HST+WFC3 \citep{2013ApJ...779...34C}, sensitive to $H\alpha$ in the range $0.7<z<1.5$ up to faint flux levels ($3-5 \times 10^{-17}$ erg s$^{-1}$ cm$^{-2}$) on a small area ($\sim 0.037$ deg$^2$); and the grism survey with HST+NICMOS by \citet{2009ApJ...696..785S}, on $\sim$ 104 arcmin$^2$ over the redshift range $0.7<z<1.9$. 
To extend the models to a broader redshift range and better constrain the evolutionary form, we include other luminosity functions available at lower redshifts.
Different subsets of input data are adopted in the three models to describe the evolution in redshift of the \HaLF, as well as to explore the robustness of the predictions.

\begin{table*}
\caption{\label{tab:LFs} Empirical Schechter parameters for the various surveys considered, ordered by redshift.
Units are Mpc$^{-3}$ ($\phi_\star$), erg s$^{-1}$ ($L_\star$) and deg$^2$ (Area).}
\small{
\begin{tabular}{lllllllll}
\hline\hline
\multicolumn{9}{l} { }
\\[-1ex]
 Redshift & ~~~~$\alpha$ & $\log_{10}L_\star$ & $\log_{10}\phi_\star$ & delta-$z$ & Area & Instr. & Reference(s) & Models \\[1ex]
\hline
     0.0225  &      $-1.3$   &    41.47   &    $-2.78$ & 0-0.045 & 471 & prism & \citet{1995ApJ...455L...1G}  & 1,2 \\ %
 0.07, 0.09  &     $-1.59$   &    41.65   &    $-3.14$ & 0.02 & 0.24 & Narrow-band & \citet{2007ApJ...657..738L}       & 1,2  \\ %
        0.2  &     $-1.35$   &    41.52   &    $-2.56$ & 0-0.3 & 0.03 & CFHT & \citet{1998ApJ...495..691T} & 1,2  \\ %
       0.24  &     $-1.35$   &    41.54   &    $-2.65$ & 0.02 & 1.54 & Narrow-band &  \citet{2008ApJS..175..128S}   & 1,2  \\ %
       0.24  &     $-1.70$   &    41.25   &    $-2.98$ & 0.02 & 0.24 & Narrow-band &  \citet{2007ApJ...657..738L}   & 1,2  \\ %
\hline
        0.4  &     $-1.28$   &    41.29   &     $-2.4$ & 0.02 & 0.24 & Narrow-band &  \citet{2007ApJ...657..738L}   & 1,2      \\ %
        0.4  &     $-1.75$   &    41.57   &    $-3.12$ & 0.02 & 2 & Narrow-band &  HiZELS \citep{2013MNRAS.428.1128S}\textsuperscript{e} & 1,2,3  \\ %
        0.6  &     $-1.27$   &    41.72   &    $-2.51$ & 0.3-0.9 & 0.037 &  HST+WF3 &  WISP \citep{2013ApJ...779...34C} & 1,2,3  \\ %
\hline
       0.73  &     $-1.31$   &    41.97   &   $-2.319$ & 0.5-1.1 & 0.031 & ISAAC &  \citet{2002MNRAS.337..369T}  & 1,2  \\ %
       0.84  &     $-1.56$   &    41.92   &    $-2.47$ & 0.04 & 2 & Narrow-band &  HiZELS \citep{2013MNRAS.428.1128S}\textsuperscript{e} & 1,2,3   \\ %
\hline
       1.05  &     $-1.39$   &    42.49   &   $-2.948$ & 0.7-1.4 & 0.029 & HST+NICMOS &  \citet{2009ApJ...696..785S}     & 1,2,3    \\ %
        1.2  &     $-1.43$   &    42.18   &     $-2.7$ & 0.9-1.5 & 0.037 & HST+WF3 &  WISP \citep{2013ApJ...779...34C} & 1,2,3  \\ %
       1.25  &      $-1.6$   &    42.87   &    $-3.11$ & 0.7-1.8 & 0.0012 & HST+NICMOS &  \citet{2000AJ....120.2843H}   & 1,2    \\ %
        1.3  &     $-1.35$   &    42.81   &   $-2.801$ & 0.7-1.9 & 0.018 & HST+NICMOS &  \citet{1999ApJ...519L..47Y} & 1,2,3    \\ %
\hline
       1.47  &     $-1.62$   &    42.23   &    $-2.61$ & 0.04 & 2 & Narrow-band & HiZELS \citep{2013MNRAS.428.1128S}\textsuperscript{e} & 1,2,3   \\ %
       1.65  &     $-1.39$   &    42.55   &   $-2.768$ &0.7-1.9 & 0.029 & HST+NICMOS &  \citet{2009ApJ...696..785S}\textsuperscript{c}      & 1,2,3      \\ %
\hline
       2.23  &     $-1.59$   &    42.53   &    $-2.78$ & 0.04 & 2 & Narrow-band & HiZELS \citep{2013MNRAS.428.1128S}\textsuperscript{a} & 1,2,3  \\ %
       2.23  &     $-1.72$   &    43.22   &    $-3.96$ & 0.04 & GOODS-S & Narrow-band &  {\citet{2010A&A...509L...5H}}\textsuperscript{b}  & 1,2    \\ %
       2.23  &      $-1.6$   &    43.07   &    $-3.45$ & ~ & ~ & ~ &  {\citet{2010A&A...509L...5H}}\textsuperscript{c} & 1,2 \\ %
       2.23  &     $-1.35$   &    42.83   &     $-3.2$ & 0.04 & 0.6 & Narrow-band &  HiZELS \citep{2008MNRAS.388.1473G}\textsuperscript{a,d} & 1,2    \\ %
\hline\hline
\multicolumn9l{\textsuperscript{a}\footnotesize{The \citet{2013MNRAS.428.1128S} analysis includes a superset of the fields used for the earlier HIZELS paper \citep{2008MNRAS.388.1473G}.}} \\
\multicolumn9l{\textsuperscript{b}\footnotesize{{\citet{2010A&A...509L...5H}} results from their internal HAWK-I data}} \\
\multicolumn9l{\textsuperscript{c}\footnotesize{{\citet{2010A&A...509L...5H}} results from a joint fit including their internal HAWK-I data and the \citet{2008MNRAS.388.1473G} data.}} \\
\multicolumn9l{\textsuperscript{d}\footnotesize{In the original luminosity function the $\phi_\star$ parameter quoted contains the conversion factor of $\ln 10$ (private communications by authors).}} \\
\multicolumn9l{\textsuperscript{e}\footnotesize{We applied an aperture correction 
of $+0.02$, $+0.07$, $+0.07$, and $+0.06$ dex at $z=0.40$, 0.84, 1.47, and 2.23, respectively.}
}
\end{tabular}
}
\end{table*}

Our focus is on predictions for the yield of galaxy redshift surveys, so we work in terms of observed \Ha\ flux, i.e. with no correction for extinction in the target galaxy.
Generally, in \Ha\ surveys direct measurements of extinction are unavailable, and thus
require purely statistical corrections. Usually an average extinction of 1 mag. has been adopted
by most of the authors (see \citet{Hopkins2004}, \citet{2013MNRAS.428.1128S}).
In cases where such corrections have been applied in the literature, we have undone the correction.
Furthermore, in many of the input data sets, \Ha\ is partially or fully blended with the \nii\ doublet, and the inference of separate \Ha\ and \nii\ fluxes is based on different assumption on their ratio or on different scaling relation.
The luminosity functions presented here are for \Ha\ only, \emph{not} \Ha+\nii,
since future space missions, like {\slshape Euclid} and {\slshape WFIRST-AFTA}, 
will have higher spectral resolution than HST and will be partially able to resolve the \Ha+\nii\ complex.
The inferred \Ha\ luminosity function is sensitive to the prescription for the \nii\ correction.
In Appendix~\ref{app:M3}
we explore the effects of different treatments.

In Table \ref{tab:LFs} we list the compilation of \HaLF\ Schechter parameters provided by various \Ha\ surveys spanning the redshift range $0 < z < 2$ which are used in this work. We also list the subset of data used in each model. Schechter parameters have been converted to the same cosmology and to the original \Ha\ extincted luminosity, when necessary. 
The luminosities in HiZELS LFs have been further corrected for aperture. These corrections are based on the fraction of a Kolmogorov seeing disk of the specified size (0.9, 0.8, 0.8, or 0.8 arcsec, full width at half maximum, specified by \citealt{2013MNRAS.428.1128S}) convolved with an exponential profile disk of half-light radius 0.3 arcsec. Similar correction has been adopted by \citet{Sobral15} within the same survey (see their Section 2.4). Variations of this procedure are explored in Appendix~\ref{app:M3}, where we find a $10$\%\ change in the abundance of \Ha\ emitters in the range relevant for {\slshape Euclid} if this correction is turned off entirely, and a 2\%\ change if the measured size-flux relation \citep{2013ApJ...779...34C} is used in place of a single reference value.

\begin{figure*}
\includegraphics[angle=270,width=180mm]{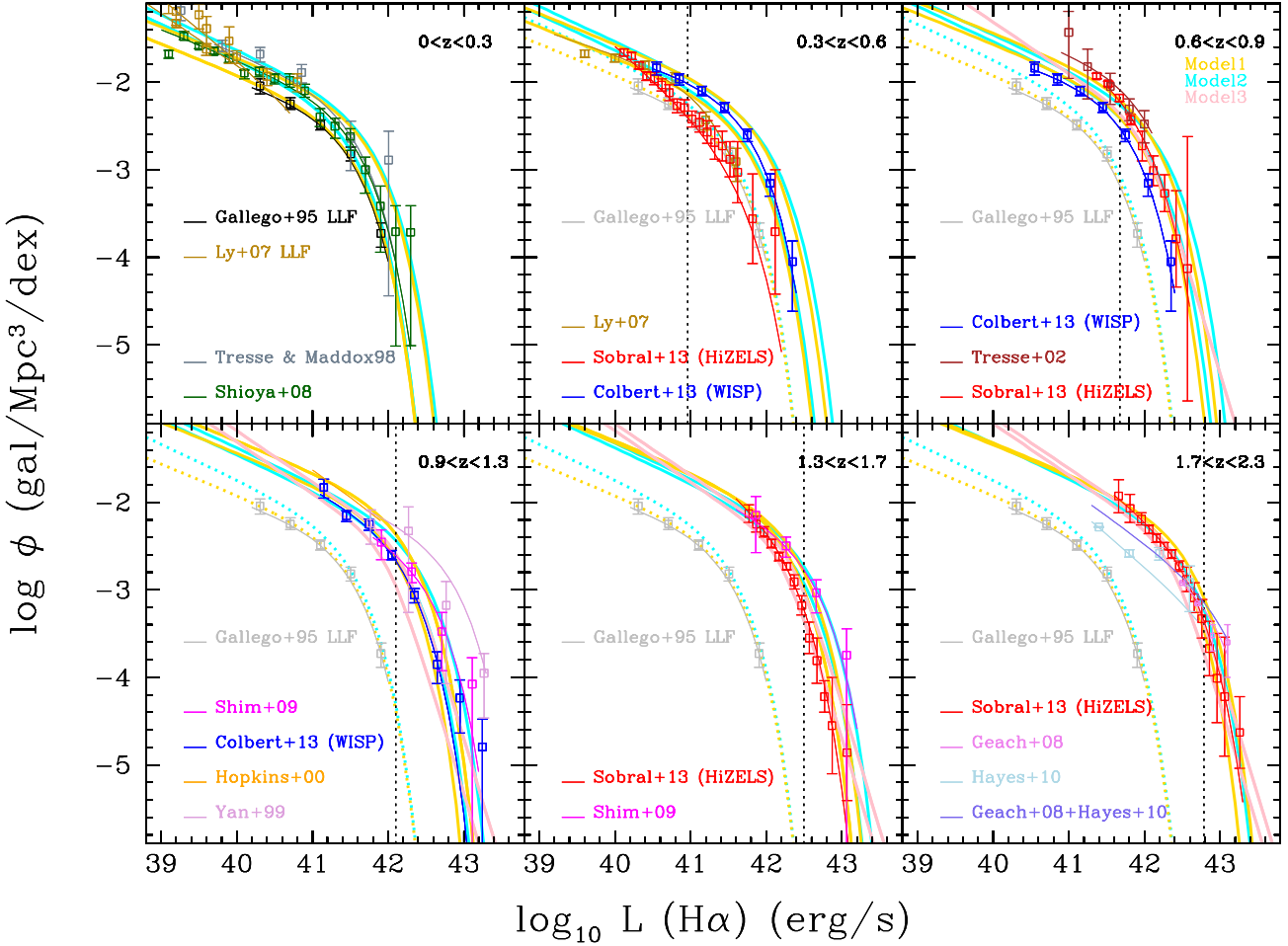}
\caption{\label{fig:haLF} H$\alpha$ LFs at various redshifts.
The dotted lines mark the nominal flux limit of {\slshape Euclid} ($3\times 10^{-16}$ erg cm$^{-2}$ s$^{-1}$) in the lower bound of each redshift range. 
Observed Schechter LFs are shown as thin lines and squares in the observed luminosity range and listed in the labels. 
For comparison, the LFs from Empirical Models 1, 2, and 3 are shown (in yellow, cyan, and pink, respectively) as thick lines in the same redshift range (shown in the two extremes of each redshift bin).
}
\end{figure*}

Figure \ref{fig:haLF} shows the empirical \HaLF s  analysed and used in our models, divided into several redshift bins from $z=0$ (reported in all panels) to $z=2.3$. 
For clarity the Schechter fits and data have been plotted only in the range of luminosities covered from each survey; the Schechter parameters have been also shown as a function of redshift (Figure \ref{fig:LFP}).
Besides the \HaLF s listed in Table \ref{tab:LFs} and shown in Figure \ref{fig:haLF}, we have compared fewer additional \HaLF s available \citep[] {Gunawardhana13, 2012PASP..124..782L, 2011ApJ...726..109L}, finding that they are consistent with the data used in this work. 
In the highest redshift bin analysed we further compare the observed \Ha\ LFs with the ones derived indirectly from UV in a sample of LBGs at $1.7<z<2.7$ \citep{2008ApJS..175...48R}, finding it slightly higher than the direct observed \Ha\ LFs.
Very recently new \HaLF s have become available at high redshift using larger area than before, both from narrow-band imaging survey \citep[CF-HiZELS,][]{Sobral15} and using slitless spectroscopy from the analysis of a wider portion of the WISP survey \citep{Mehta15}. Both the new \HaLF s are consistent with previous determinations. In the following section we compare our models to these new data. We have not attempted instead to include in our models these new, but not independent, 
LFs determinations which does not reduce cosmic variance substantially and in the case of 
\citet{Mehta15} and \citet{2008ApJS..175...48R} have been derived indirectly from \oiii\ lines and UV fluxes, respectively 
 at high-$z$.

From the data analysed in this work, we note that in the local Universe the shape of the \HaLF\ is well established and
characterized across a large range of luminosities
\citep{1995ApJ...455L...1G,2007ApJ...657..738L}.
Over the past decade, improvements in NIR grisms, slit spectroscopy, and narrow band surveys have allowed 
the evolution of the LF to be tracked out to $z\simeq 2$ (see references in Table \ref{tab:LFs}),
not only at the bright end but also below the characteristic $L_\star$.
However, we note that at $z>0.9$ the various empirical \HaLF s start to disagree, as confirmed by their Schechter parameters.
Despite the empirical uncertainties it is clearly evident the  strong luminosity 
evolution of the bright end of the \HaLF\ with increasing redshift, also confirmed by the evolution of $L_\star$ by about an order of magnitudes over the whole redshift range. On the other hand, the amount of density evolution is still not completely clear, as well as the exact value and evolution of the faint end slope, as attested by the evolution with redshift of the $\phi_\star$ and $\alpha$ Schechter parameters.

\section {Modelling the \Ha\ luminosity function evolution}
\label{sec:models}

\begin{figure}
\includegraphics[angle=0,width=88mm]{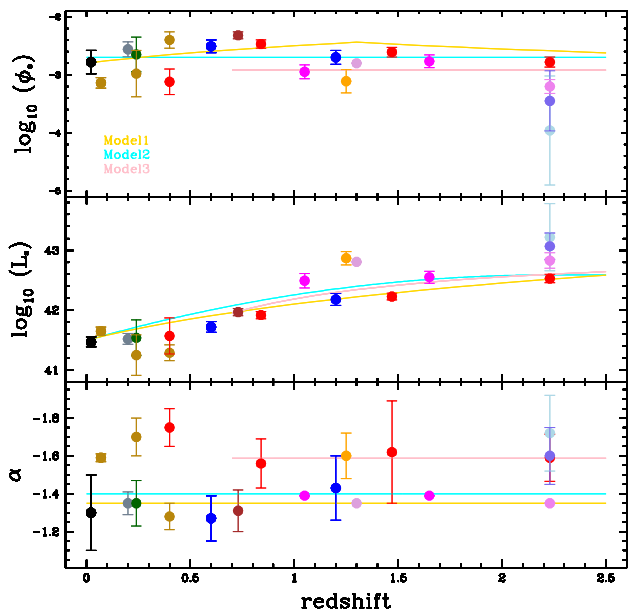}
\caption{\label{fig:LFP} 
H$\alpha$ LFs empirical Schechter parameters 
(using the same colours as Figure \ref{fig:haLF}) as a function of redshift (at the center redshift of each surveys), along with the evolution of parameters in the models.
}
\end{figure}

As outlined in the previous section and shown in Figs. \ref{fig:haLF} and \ref{fig:LFP}, in the relevant redshift range for future \Ha\ missions, existing \Ha\ LF  measurements show large  uncertainties and are often inconsistent with one another.
In light of these uncertainties, we cannot recommend a unique model with only its statistical error associated, because this would be based on a predefined evolutionary and luminosity function shape. We, rather, present three models based on different treatments of the input data (named “Model 1”, “Model 2” and “Model 3”, hereafter). 
In particular we adopt three different evolutionary forms 
to describe the uncertain evolution of the \HaLF.
For the shape of the luminosity function, three functional forms were considered. 
The simplest is the Schechter function.
We also adopt, different methodologies, subsets of input data, 
and treatment of systematic errors to explore the uncertainties and robustness of the predictions.

\subsection{Model 1}

In this model we used a \citet{1976ApJ...203..297S} parametrization for the luminosity functions and
an evolutionary form similar to \citet{2010MNRAS.402.1330G},
\begin{equation}
  \phi(L,z) \, \D L = \phi_\star \left(
  \frac{L}{L_\star} \right)^{\alpha} \rme^{-L/L_\star} \,
  \frac{ \D L }{ L_\star },
\label{eqn:schechter}
\end{equation}
where
\begin{list}{$\bullet$}{}
\item $\phi_\star$ is the characteristic density of H$\alpha$ emitters;
\item $\alpha$ is the faint-end slope;
\item $L_\star$ is the characteristic luminosity at which the H$\alpha$ luminosity function falls by a factor of $\rme$ from the extrapolated faint-end power law. It has a value at $z=0$ of $L_{\star,0}$;
\item  and $\rme=2.718...$ is the natural logarithm base.
\end{list}

We adopt the same evolutionary form for $L_\star$ assumed in \citet{2010MNRAS.402.1330G}, and introduce an evolution in $\phi_\star$,
\begin{equation}
L_{\star,z} = L_{\star,0}  (1 + z)^{\delta }
\end{equation}
and
\begin{equation}
\phi_{\star,z}= \left\{ 
\begin{array}{ccc}
\phi_{\star,0}  (1+z)^\epsilon & & z<z_{\rm break} \\
\phi_{\star,0} (1+z_{\rm break})^{2\epsilon} (1+z)^{-\epsilon} & & z>z_{\rm break}
\end{array}
\right.;
\end{equation}
thus $\phi_{\star,0}$ is the characteristic number density today, which is taken to scale as $\propto (1+z)^\epsilon$ at $0<z<z_{\rm. break}$ and $\propto (1+z)^{-\epsilon}$ for $z>z_{\rm break}$.

Because the Schechter parameters are correlated, 
we do not rely on the evolution of the empirical Schechter parameters to constrain their evolution since 
a unique or fixed $\alpha$ value has not been found or fixed.
We instead attempt to reproduce, by mean of a $\chi^2$ approach, 
the observed luminosity functions at different luminosities and redshifts, as described by their Schechter functions in the 
luminosity range covered by the observations. 
We, therefore, find the best parameters (reported in Table \ref{tab:models}) 
$\alpha= -1.35$, $L_{\star,0}=10^{41.5}$ erg s$^{-1}$, $\phi_{\star,0}=10^{-2.8}$ Mpc$^{-3}$, $\delta=2$, $\epsilon=1$, and $z_{\rm break}=1.3$.

\subsection{Model 2}

We adopt the same Schechter function for the LFs as for Model 1, 
but change the evolutionary form for $L_\star$ as follows:
\begin{equation}
\log_{10} L_{\star,z} = -c(z-z_{\rm break})^2 + \log_{10} L_{\star,z_{\rm break}}.
\end{equation}
In this model we normalize the evolution of $L_\star$ to the maximum redshift available 
($z_{\rm break}$=2.23) and we assume no evolution for $\phi_{\star}$, i.e. $\epsilon=0$.
Using the same fitting method used for Model 1, to reproduce the observed luminosity functions at different 
redshift, we find the best fit parameters (reported in Table \ref{tab:models})
$\alpha= -1.4$, $\phi_{\star,0}=10^{-2.7}$ Mpc$^{-3}$, 
$c=0.22$, $L_{\star,z_{\rm break}}=10^{42.59}$ erg s$^{-1}$
($\epsilon=0$, $z_{\rm break}=2.23$).

\subsection{Model 3}

Model 3 is a combined fit to the HiZELS \citep{2013MNRAS.428.1128S}, WISP \citep{2013ApJ...779...34C}, and NICMOS \citep{1999ApJ...519L..47Y, 2009ApJ...696..785S} data only.
The procedure was designed specifically for use only in the redshift ranges under consideration for the {\slshape Euclid} and {\slshape WFIRST-AFTA} slitless surveys (in particular at $0.7<z<2.23$). As such, 
only the three largest compilations in the relevant redshift and flux range were used; 
in particular the low-$z$ data is not part of the Model 3 fit and we do not display Model 3 results at $z<0.6$. 
We obtained the Model 3 luminosity function parameters and their uncertainties using a Monte Carlo Markov chain (MCMC). In Appendix~\ref{app:M3}, we explore the dependence of the Model 3 fits on the assumptions, fitting methodology, and subsets of the input data used, which is a useful way to assess some types of systematic error.

Model 3 has the advantage of being fit directly to luminosity function data points, and not to the analytic Schechter fit, 
as done in Model 1 and 2.
However, we do not recommend its use at $z\lesssim 0.6$ since it does not incorporate the low-redshift data.

A fit to the data also requires a likelihood function (or error model), in addition to central values for the data points. The luminosity functions provided by individual groups contain the Poisson error contribution, as well as estimated errors from other sources (e.g. uncertainties in the completeness correction). The construction of this model is complicated by two issues: cosmic variance and asymmetric error bars.
The model for cosmic variance uncertainties is described in Appendix~\ref{app:cv}. Our default fits are performed using the CV2 model, which allows for a luminosity-dependent bias. The alternative models are CV1, which assumes a luminosity-independent bias, and a no-cosmic variance model.
Fitting large numbers of data points can lead to statistically significant biases if the error bars are asymmetric and this is not properly accounted for in the analysis. 
The treatment of asymmetric error bars is discussed in Appendix~\ref{app:Poisson}. We use the Poisson option for our primary fits, and consider the other variations in Appendix~\ref{app:M3}.

Several models for evolving luminosity functions were investigated and model parameters were fit using the MCMC. All models require a break in the luminosity function to describe the data; the break position $L_\star$ is taken to be a function of redshift,
\begin{equation}
\log_{10} L_{\star,z} = \log_{10} L_{\star,\infty} + \left( \frac{1.5}{1+z} \right)^\beta \log_{10} \frac{L_{\star,0.5}}{L_{\star,\infty}} .
\label{eq:model1-a}
\end{equation}
Where necessary, we write $L_{\star,z}$ to denote the characteristic luminosity at redshift $z$. For the shape of the luminosity function, three forms were considered. The simplest and most commonly used is the Schechter function,
but the exponential cutoff is a poor fit to the observations.
Two alternative models were considered to fix this: a hybrid model
\begin{equation}
\phi(L,z) =  \frac{\phi_\star}{L_\star} \left( \frac L{L_\star} \right)^\alpha {\rm e}^{-(1-\gamma)L/L_\star}
\left[ 1 + (\rme-1)\left(\frac{L}{L_\star}\right)^2 \right]^{-\gamma}
~~~({\rm hybrid}),
\label{eq:model1-b.hybrid}
\end{equation}
that mixes broken power law and Schechter behaviour; and a broken power law,
\begin{equation}
\phi(L,z) = \frac{\phi_\star}{L_\star} \left( \frac L{L_\star} \right)^\alpha
\left[ 1 + (\rme-1)\left(\frac{L}{L_\star}\right)^\Delta \right]^{-1}
~~~({\rm broken~power~law}).
\label{eq:model1-b.broken}
\end{equation}
The broken power law is the simplest function\footnote{The factor of ${\rm e}-1=1.718...$ in Eq.~(\ref{eq:model1-b.broken}) does not lead to any physical change in the model -- it is equivalent to a re-scaling of the break luminosity $L_\star$. With the stated normalization, $L_\star$ is the luminosity at which the LF falls to 1/e of the faint-end power law, which is the same meaning that it has in the Schechter function; without this factor, $L_\star$ would correspond to the luminosity at which the LF falls to 1/2 of the faint-end power law.} that interpolates between a faint-end power law $\phi\propto L^\alpha$ and a bright-end power law $\phi\propto L^{\alpha+\Delta}$. Both of these models are empirically motivated: they were introduced to fit the shallower (than Schechter) cutoff at high $L$. 
We also tried an alternative functional form used for low-redshift FIR and H$\alpha$ data (\citealt{Saunders90}, Eq. 1; \citealt{Gunawardhana13}, Eq. 11), however the fit is worse than for the broken power law ($\chi^2$ is higher by 8.1, with the same number of degrees of freedom) so we did not adopt it.

The Schechter function has one fewer parameter than the others, so in this case $\phi_\star$ was allowed to have an exponential evolution with scale factor $a=1/(1+z)$,
\begin{equation}
\log_{10}\phi_{\star,z} = \log_{10}\phi_{\star,1} + \frac{\D\log_{10}\phi_\star}{\D a}\left( \frac1{1+z} - \frac12 \right),
\label{eq:sDE}
\end{equation}
with $(\D/\D a)\log_{10}\phi_\star$ taken to be a constant. The broken power law model was used for the reference fit, since it gives the best $\chi^2$.

These models have $N_{\rm par} = 6$ parameters, 3 parameters besides the standard Schechter parameters ($\phi_\star$, $\alpha$, $L_\star$),  whose meaning, is as follows:
\begin{list}{$\bullet$}{}
\item $(\D/\D a)\log_{10}\phi_\star$ (for the Schechter function) characterizes density evolution; it is positive if H$\alpha$ emitters get more abundant at late times.
\item $\gamma$ (hybrid model only) interpolates between an exponential or Schechter-like cutoff at high $L$ ($\gamma=0$) or a broken power law form ($\gamma=1$: the power law index changes by 2 between the faint and bright ends). Values of $\gamma>1$ are not allowed. 
\item $\Delta$ (broken power law model only) is the difference between bright and faint-end slopes.
\item $L_{\star,z}$ 
has a high-$z$ extrapolated limiting value\footnote{Of course, at very high redshift the \Ha LF must fall off since there are no galaxies. We remind the reader that the empirical models built here may not be valid outside the range of redshifts spanned by the input data.} ($L_{\star,\infty}$) and a value at $z=0.5$ ($L_{\star,0.5}$). The sharpness of the fall-off in $L_\star$ at low redshift is controlled by $\beta$. In some models, $\log_{10}L_{\star,2.0}$ (the value at $z=2.0$) was used instead of $\log_{10}L_{\star,\infty}$ to reduce the degeneracy with $\beta$.
\end{list}
The three models differ most strongly in their assumed form at the high-luminosity end: the broken power law has a power law scaling (with slope $\alpha-\Delta$), whereas the Schechter function has an exponential cutoff. The hybrid model has an exponential cutoff if $\gamma<1$, but its steepness is decoupled from $L_\star$ -- as $\gamma\rightarrow 1$, the scale luminosity in the cutoff $L_\star/(1-\gamma)$ can be much greater than the luminosity at the break $L_\star$.

The likelihood evaluation predicts the luminosity function averaged over a bin of $\log_{10}L_{{\rm H}\alpha}$ and enclosed volume $[D(z)]^3$ using $N_{\rm G}\times N_{\rm G}$ Gauss-Legendre quadrature scheme (for slitless surveys) or an $N_{\rm G}$-point Gaussian quadrature (for narrow-band surveys, where there is no need to do a redshift average). The fiducial value of the quadrature parameter is $N_{\rm G}=3$.
A flat prior was used on the 5 or 6 parameters $(\alpha,\log_{10}\phi_\star,\log_{10}L_{\star,2.0},\log_{10}L_{\star,0.5},\beta$ and $\gamma$ or $\Delta$, as appropriate). Chains are run with a Metropolis-Hastings algorithm; at the end a minimizing algorithm is run on the $\chi^2$ to find the maximum likelihood model.

\begin{figure}
\includegraphics[width=88mm]{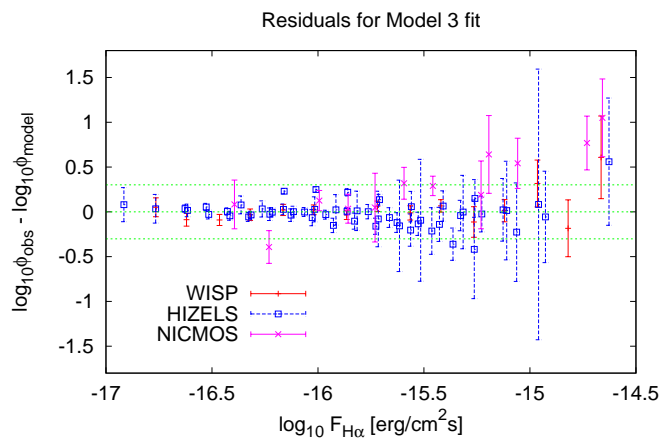}
\caption{\label{fig:residuals} Residuals to the H$\alpha$ luminosity function fits for Model 3, plotted as a function of observed-frame H$\alpha$ flux at the bin centre (horizontal axis). All redshifts are plotted together. The green lines show the fit line and factors of 2 above and below. The error bars shown do \emph{not} include the cosmic variance, which is included in the fit but is highly correlated across luminosity bins.}
\end{figure}

The best fit model (lowest $\chi^2$) is the broken power law model with parameters 
(reported in Table \ref{tab:models})
$\alpha = -1.587$, $\Delta = 2.288$, $\log_{10}\phi_{\star,0} = -2.920$, $\log_{10} L_{\star,2.0} = 42.557$, $\log_{10} L_{\star,0.5} = 41.733$, and $\beta=1.615$ (and the corresponding $\log_{10} L_{\star,\infty} = 42.956$).
The faint-end slope of $\alpha=-1.587$ is $2.0\sigma$ shallower than the ultraviolet LF slope of $-1.84\pm0.11$ measured by \citet{2008ApJS..175...48R}, and $1.5\sigma$ shallower than $-1.73\pm0.07$ measured by \citet{2009ApJ...692..778R} at $z\approx 2$.
The residuals from this fit are shown in Figure~\ref{fig:residuals}.

\begin{table*}
\caption{\label{tab:models} Fit parameters for the three models considered. 
Best fit central values and 2$\sigma$ errors (without uncertainties when fixed).
Units are Mpc$^{-3}$ ($\phi_\star$) and erg s$^{-1}$ ($L_\star$).} 
\small{
\begin{tabular}{lrrrrrr}
\hline\hline\multicolumn{7}{c} { }
\\[-1ex]
& $\alpha$ & $\log_{10}\phi_{\star,0}$ & $\log_{10}L_{\star,0}$ & $\delta$ & $\epsilon$ & $z_{break}$ \\[1ex]
{\tt Model 1} & $-1.35^{+0.10}_{-0.15}$ & $-2.80^{+0.15}_{-0.18}$ & $41.50^{+0.11}_{-0.11}$ & $2.0^{+0.1}_{-0.1}$ & $ 1.0^{+0.1}_{-0.1}$ & $ 1.3^{+0.1}_{-0.1}$ \\[1ex]
\hline
& $\alpha$ & $\log_{10}\phi_{\star,0}$ & $\log_{10}L_{\star,z_{break}}$ & $c$ & $\epsilon$ & $z_{break}$ \\[1ex]
{\tt Model 2} & $-1.40^{+0.10}_{-0.15}$ & $-2.70^{+0.17}_{-0.17}$ & $42.59^{+0.10}_{-0.12}$ & $0.22^{+0.05}_{-0.05}$ & $ 0.0$ & $ 2.23$ \\[1ex]
\hline
& $\alpha$ & $\log_{10}\phi_{\star,0}$ & $\log_{10}L_{\star,2.0}$ & $\log_{10}L_{\star,0.5}$ & $\Delta$ & $\beta$ \\[1ex]
{\tt Model 3} & $-1.587^{+0.132}_{-0.119}$ & $-2.920^{+0.183}_{-0.175}$ & $42.557^{+0.109}_{-0.119}$ & $41.733^{+0.150}_{-0.142}$ & $ 2.288^{+0.410}_{-0.379}$ & $ 1.615^{+0.947}_{-1.196}$ \\[1ex]
\hline\hline
\end{tabular}
}
\end{table*}

\section{Comparison to observed luminosity functions}
\label{sec:compare-e}

The three empirical models constructed are plotted in Figure~\ref{fig:haLF} in different redshift bins, 
and compared to the observed H$\alpha$LFs.
The Schechter parameters for data and models are also shown, only for illustrative purpose, in Figure \ref{fig:LFP}. Note, however, that, since the parameters are correlated, a direct comparison between them is not straightforward, in particular Model 3 assumes also a different form for the LFs.

\begin{figure*}
\center{\includegraphics[angle=270,width=180mm]{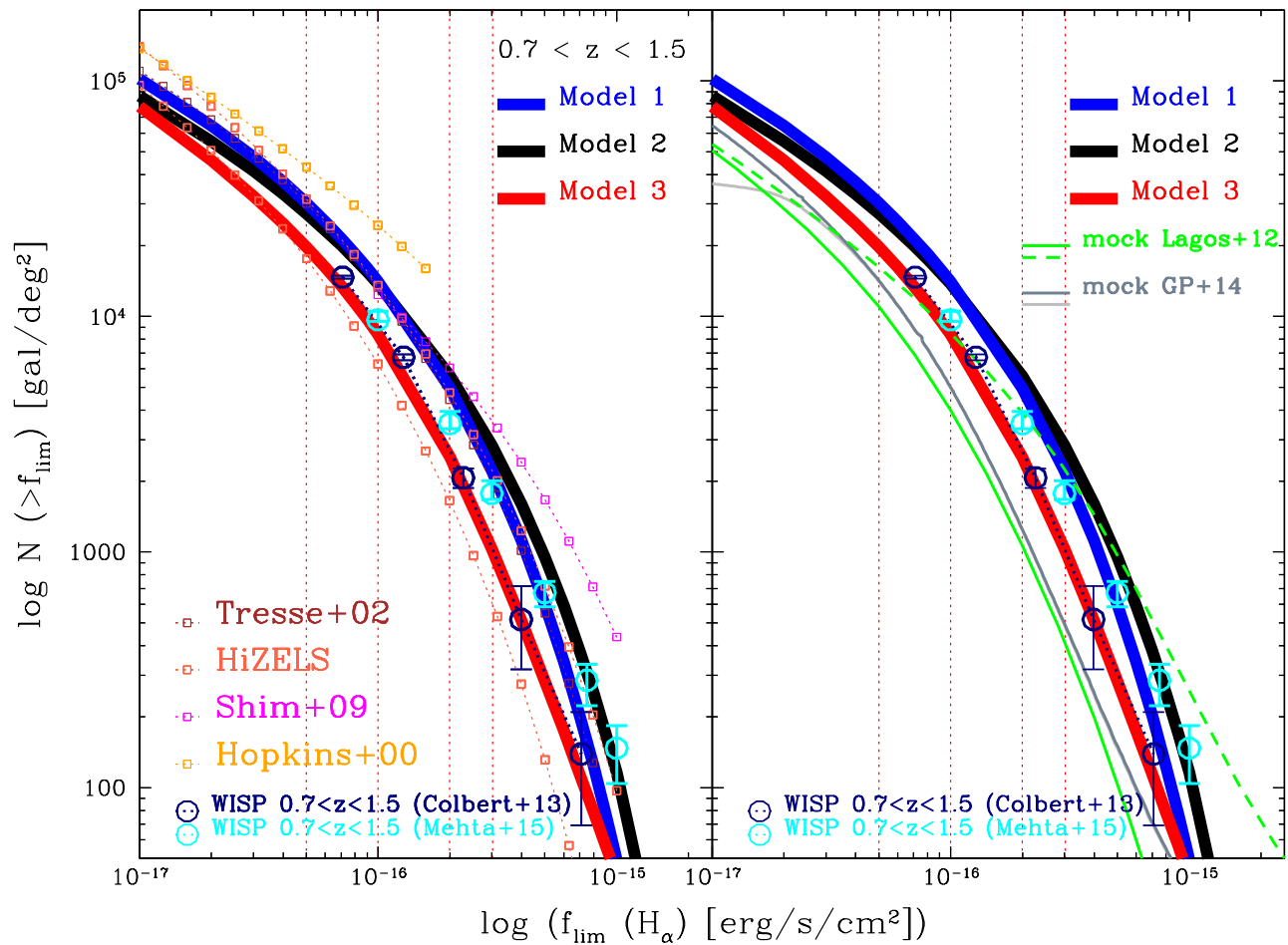}}
\caption{\label{fig:hacounts} {\it Left panel:} Cumulative H$\alpha$ number counts, integrated over the redshift ranges $0.7<z<1.5$ (WISP range). 
The observed counts from the WISP survey \citep{2013ApJ...779...34C} are shown (blue circles) and from new WISP analysis by \citet{Mehta15} (cyan circles), and compared to the empirical Model 1, 2, and 3, (blue, black and red lines, respectively). Also shown (as dotted lines and empty squares) are the counts obtained integrating the observed LFs (see legend) in the same redshift range. {\it Right panel:} The same cumulative \Ha\ number counts compared to the predictions from L12 mocks (green dashed and solid lines using intrinsic and extincted \Ha\ fluxes, respectively) and GP14 mocks (dark and light grey for $H<27$ and $H<24$ mocks, respectively).
}
\end{figure*}

Given the large scatter in the observed LFs covering similar redshift ranges, all the 3 models provide a reasonable
description of the data.
Indeed, while it is difficult to choose a best model, among the three, 
overall they describe well the uncertainties and the scatter between different observed LFs, 
in particular at high-$z$.

Comparing different redshift bins, it is evident that at low-$z$ the models evolve rapidly in luminosity, as
clearly visible also in the evolution of $L_{\star}$ parameter, resulting in an increase of the density of 
high luminosity H$\alpha$ emitters. Since at high-$z$, instead, all 3 models evolve mildly 
in luminosity and density, or even slightly decrease in density (for Model 1), as a consequence the density 
of high-$L$ objects is almost constant.

Finally, we note that the main difference between the 3 models at all redshifts is at the bright-end of the luminosity function. Model 1 has the lower high-luminosity end, but is similar in shape to Model 2, (both
assuming a Schechter form), while Model 3 has the most extended bright-end, while it is slightly lower 
at intermediate luminosities, and has the steepest faint-end slope. This occurs because of the different 
functional form used in Model 3. 
Current uncertainties in the bright-end of the empirical \HaLF s, do not allow strong constraint on the functional form.
Actually, recent analysis of GAMA and SDSS surveys \citep{Gunawardhana13, Gunawardhana15} and of WISP \citep{Mehta15} suggest a LF more extended than a Schechter function but only at very bright luminosity ($>10^{43}$ erg/s). Further analysis on wider area will provide new insight on this issue.


\section{Number counts and redshift distribution of \Ha\ emitters}
\label{sec:abundance}

The cumulative counts, as a function of \Ha\ flux limit, predicted by the models are shown in Figure~\ref{fig:hacounts}. 
We derive the cumulative counts in the same redshift range covered by the WISP slitless data, 
i.e. $0.7<z<1.5$.  For comparison we show the observed \Ha\ WISP counts, taken from Table 2 
of \citet{2013ApJ...779...34C}
and corrected for \nii\ emission as indicated in the original paper, 
with $L_{{\rm H}\alpha}= 0.71(L_{{\rm H}\alpha}+L_{\nii})$, for consistency with the WISP \Ha\ LF used here.
Besides WISP counts, we show also the predicted counts 
using single luminosity functions at different redshifts (integrated over the same redshift range $0.7<z<1.5$). 
The three models reproduce well the scatter between the observed counts and observed luminosity functions, 
with Model 3 giving the lowest counts due to the large weight assigned to the (lower amplitude) HiZELS and WISP samples.

At the depth and redshift range of the originally planned {\slshape Euclid} Wide grism survey \citep{2011arXiv1110.3193L}, i.e  $F_{{\rm H}\alpha}> 3 \times 10^{-16}$ erg cm$^{-2}$ s$^{-1}$, and $1.1<\lambda<2.0\ \mu$m (sampling H$\alpha$ at $0.70<z<2.0$), Models 1, 2, and 3 predict about 2490, 3370, and 1220
\Ha\ emitters per deg$^{2}$, respectively. 
This increases to 42500, 39700, and 28100 \Ha\ emitters per deg$^{2}$
for $F_{{\rm H}\alpha} > 5 \times 10^{-17}$ erg cm$^{-2}$ s$^{-1}$ as originally planned for the Deep {\slshape Euclid} specroscopic survey (see Table \ref{tab:dNdz}).

The {\slshape WFIRST-AFTA} mission will have less sky coverage than {\slshape Euclid} (2200 deg$^2$ instead of 15000 deg$^2$), but with its larger telescope will probe to fainter fluxes. Its grism spans the range from 1.35--1.89 $\mu$m.\footnote{The grism red limit was 1.95 $\mu$m in the original {\slshape WFIRST-AFTA} design; it was changed to 1.89 $\mu$m in fall 2014 due to an increase in the baseline telescope operating temperature.} 
The single line flux limit\footnote{The \Ha+\nii\ complex is partially blended at {\slshape WFIRST-AFTA} resolution; the exposure time calculator \citep{2012arXiv1204.5151H} now contains a correction for this effect.} 
varies with wavelength and galaxy size; at the center of the wave band for a point source, and for a pre-PSF effective radius of 0.2 arc sec (exponential profile), it is $9.5\times 10^{-17}$ erg cm$^{-2}$ s$^{-1}$. 
The three luminosity functions integrated over the {\slshape WFIRST-AFTA} sensitivity curve\footnote{D. Spergel et al., in preparation. We also used the $j=2$ galaxy size distribution in the {\slshape WFIRST} exposure time calculator \citep{2012arXiv1204.5151H}.} predict an available galaxy density of 11900, 12400, and 7200 gal deg$^{-2}$ (for Models 1, 2, and 3 respectively), in the redshift range $1.06<z<1.88$.

\begin{figure*}
\center{\includegraphics[angle=270,width=180mm]{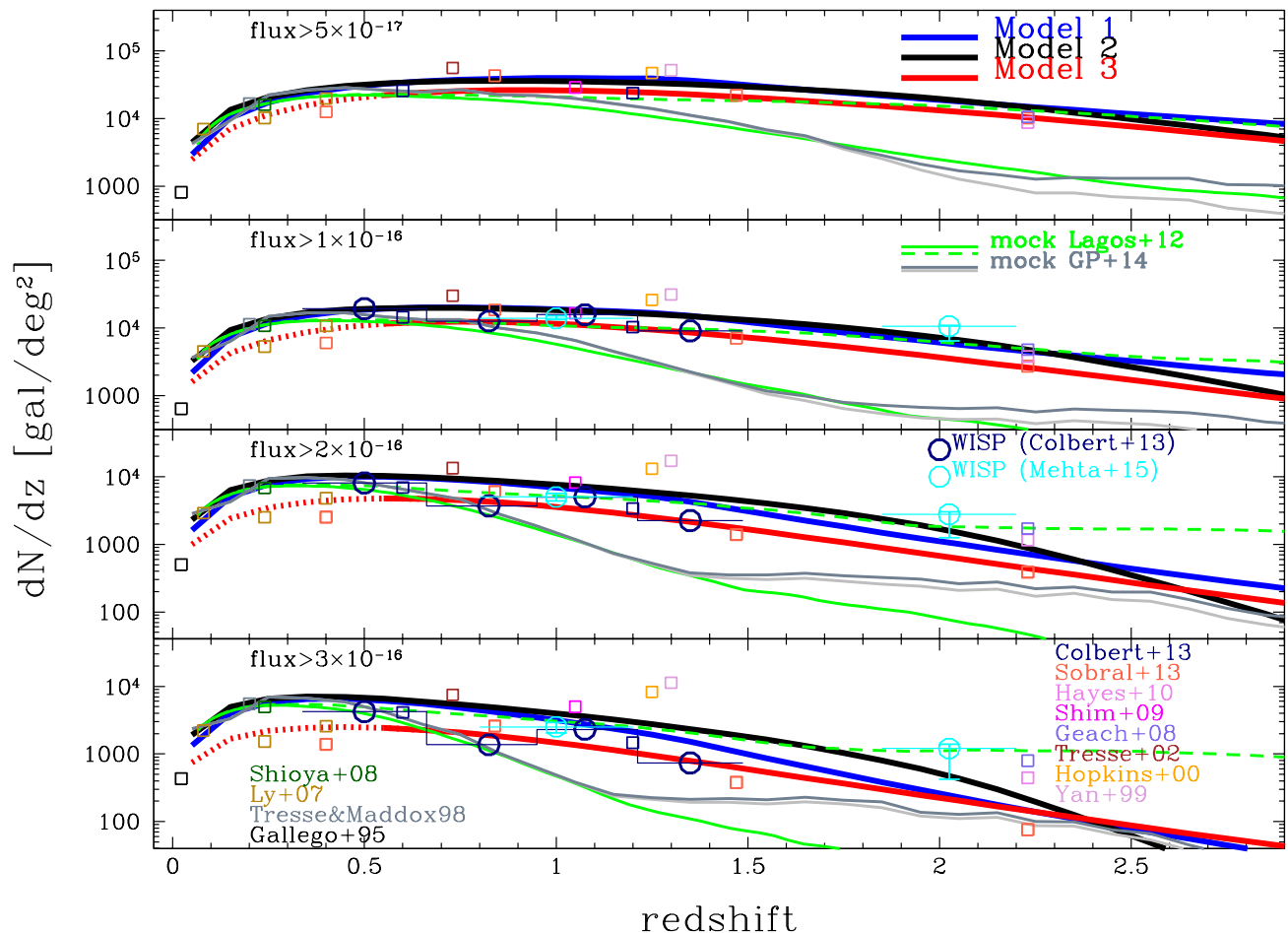}}
\caption{\label{fig:hafit} H$\alpha$ redshift distribution above various flux thresholds (from $0.5\times 10^{-16}$ erg cm$^{-2}$ s$^{-1}$ to $3\times 10^{-16}$ erg cm$^{-2}$ s$^{-1}$, from top to bottom panels). Observed redshift distributions are indicated with open circles, while data obtained integrating LFs are shown with squares. HaLF predictions from Model 1, 2, 3 are shown as thick solid lines. 
The predictions from L12 mocks (green dashed and solid lines using intrinsic and extincted \Ha\ fluxes, respectively) and GP14 mocks (dark and light grey for $H<27$ and $H<24$ mocks, respectively) are also shown.
}
\end{figure*}

The previous community standard luminosity function model, used for the 2011 {\slshape Euclid} Red Book \citep{2011arXiv1110.3193L} and in pre-2012 {\slshape WFIRST} studies \citep{Green11}, is that of \citet{2010MNRAS.402.1330G} divided by a factor of 1.257. This luminosity function predicts 7470 and 41500
\Ha\ emitters per deg$^2$ in the same redshift range ($0.7<z<2$) and flux limits ($> 3$ or $0.5 \times 10^{-16}$ erg cm$^{-2}$ s$^{-1}$) of the original wide and deep {\slshape Euclid} surveys.
This is a factor of about 2--6 
more than estimated here at bright fluxes and a similar number at faint fluxes.
The difference is partly due to the factor 
of $\ln 10 \approx 2.3$ from the convention for $\phi_\star$ in the \citet{2008MNRAS.388.1473G} luminosity function, 
and partly because \citet{2010MNRAS.402.1330G} used the brightest and highest LF by \citet{1999ApJ...519L..47Y} as the principal constraint in the $z\sim 1.3$ range; in contrast later WISP and HiZELS samples have found fewer bright H$\alpha$ emitters at this redshift.

Very recently, from the new analysis by \citet{Mehta15} of the bivariate \Ha -[OIII] luminosity function for the WISP survey, over roughly double the area used by \citet{2013ApJ...779...34C}, they expect in the range $0.7<z<2$ about 3000 galaxies/deg$^2$ for the nominal flux limit of {\slshape Euclid} ($> 3 \times 10^{-16}$ erg cm$^{-2}$ s$^{-1}$) and $\sim 20000$ galaxies/deg$^2$ for a fainter flux limit ($> 1 \times 10^{-16}$ erg cm$^{-2}$ s$^{-1}$, the baseline depth of {\slshape WFIRST-AFTA}). We note that these expectations are more consistent with our two higher models, i.e. Model 1 and 2, than with our lowest Model 3.
Figure \ref{fig:hacounts} shows their counts at $0.7<z<1.5$ (from their Table 4).

\begin{table*}
\caption{\label{tab:dNdz} Redshift distributions for a range of limiting fluxes (in units of  $10^{-16}$ erg cm$^{-2}$ s$^{-1}$) from the 3 empirical Models (1, 2, 3).
Values given are $dN/dz$ in units of deg$^{-2}$ per units redshift. 
Also listed the cumulative counts integrated over specific redshift ranges, in units of deg$^{-2}$. The predicted numbers include intrinsic extinction in the H$\alpha$ emitters and is corrected for \nii\ contamination.}
\small{
\begin{tabular}{lrrrrrrrrrrrrrrrr}
\hline\hline
\multicolumn{16}{r} { }
\\[-1ex]
 ~ & ~  & ~ & ~ & ~ & ~  & $dN/dz$ & ~ & ~ & ~  & ~ & ~ & ~ & ~ & ~ & ~ \\[1ex]
 ~ & Model 1  & ~ & ~ & ~ & ~ & Model 2  & ~ & ~ & ~ & ~ & Model 3  & ~ & ~ & ~ & ~ \\[1ex]
 Redshift & 0.5 & 1 & 2 & 3 & 5 & 0.5 & 1 & 2 & 3 & 5 & 0.5 & 1 & 2 & 3 & 5  \\[1ex]
\hline
 0.0 - 0.1  & 2924 & 2192 & 1616 & 1339 & 1044 	& 4451 & 3245 & 2329 & 1901 	& 1455 & - & - & - & - & - \\ 
 0.1 - 0.2  & 10252 & 7324 & 5078 & 4021 & 2909	& 13491 & 9406 & 6369 & 4976 	& 3543 & - & - & -  & - & -\\ 
 0.2 - 0.3  & 17381 & 11892 & 7720 & 5768 & 3773 & 20782 & 13916 & 8868 & 6572 	& 4267 & - & - & -  & - & -\\ 
 0.3 - 0.4  & 23608 & 15445 & 9287 & 6511 & 3837 & 26276 & 16921 & 10097 & 7077 	& 4190 & - & - & -  & - & -\\ 
 0.4 - 0.5  & 28730 & 17898 & 9946 & 6546 & 3462 & 30255 & 18731 & 10475 & 6964 	& 3771 & - & - & -  & - & -\\ 
 0.5 - 0.6  & 32705 & 19372 & 9964 & 6155 & 2896 & 32997 & 19659 & 10344 & 6543 	& 3243 & - & - & -  & - & -\\ 
 0.6 - 0.7  & 35612 & 20068 & 9570 & 5536 & 2297 & 34753 & 19966 & 9926 & 5987 & 2717 & 24255 & 12169 & 4739 & 2273 & 725 \\
 0.7 - 0.8  & 37594 & 20185 & 8930 & 4825 & 1757 & 35731 & 19840 & 9353 & 5388 & 2242 & 25586 & 12404 & 4517 & 2061 & 621 \\
 0.8 - 0.9  & 38813 & 19890 & 8164 & 4112 & 1310 & 36092 & 19411 & 8701 & 4794 & 1833 & 26232 & 12265 & 4181 & 1822 & 524 \\
 0.9 - 1.0  & 39423 & 19313 & 7353 & 3449 & 961 & 35961 & 18764 & 8015 & 4227 & 1487 & 26290 & 11831 & 3779 & 1579 & 437 \\ 
 1.0 - 1.1  & 39561 & 18553 & 6552 & 2861 & 698 & 35430 & 17959 & 7319 & 3697 & 1198 & 25866 & 11182 & 3350 & 1347 & 362 \\ 
 1.1 - 1.2  & 39340 & 17683 & 5794 & 2357 & 504 & 34566 & 17033 & 6627 & 3206 & 958 & 25064 & 10389 & 2923 & 1136 & 297 \\ 
 1.2 - 1.3  & 38851 & 16756 & 5097 & 1933 & 362 & 33424 & 16015 & 5946 & 2755 & 758 & 23978 & 9514 & 2518 & 949 & 243 \\ 
 1.3 - 1.4  & 36560 & 15144 & 4281 & 1515  & 250& 32045 & 14926 & 5284 & 2340 & 591 & 22691 & 8606 & 2148 & 788 & 198 \\ 
 1.4 - 1.5  & 32911 & 13107 & 3447 & 1140  & 165& 30465 & 13783 & 4642 & 1962 & 454 & 21272 & 7703 & 1817 & 652 & 162 \\ 
 1.5 - 1.6  & 29635 & 11357 & 2782 & 861  & 110	& 28714 & 12601 & 4026 & 1619 & 341 & 19779 & 6832 & 1528 & 537 & 132 \\ 
 1.6 - 1.7  & 26704 & 9856 & 2253 & 654  & 74	& 26823 & 11396 & 3440 & 1311 & 249 & 18259 & 6013 & 1279 & 442 & 107 \\ 
 1.7 - 1.8  & 24090 & 8572 & 1831 & 499  & 50	& 24820 & 10182 & 2889 & 1038 & 177 & 16749 & 5256 & 1067 & 363 & 87 \\ 
 1.8 - 1.9  & 21760 & 7471 & 1493 & 382  & 34	& 22734 & 8976 & 2378 & 801 & 121 & 15277 & 4570 & 889 & 299  & 71 \\ 
 1.9 - 2.0  & 19686 & 6527 & 1223 & 295  & 23	& 20594 & 7794 & 1912 & 599 & 79 & 13864 & 3954 & 740 & 246  & 58 \\ 
 2.0 - 2.1  & 17838 & 5716 & 1006 & 228  & 16	& 18430 & 6652 & 1496 & 432 & 49 & 12524 & 3408 & 616 & 203  & 48 \\ 
 2.1 - 2.2  & 16192 & 5019 & 830 & 178 &  11 	& 16275 & 5568 & 1134 & 298 & 28 & 11268 & 2928 & 512 & 168  & 39 \\ 
 2.2 - 2.3  & 14724 & 4419 & 689 & 140 & 7.8 	& 14169 & 4562 & 830 & 196 & 15 & 10101 & 2509 & 427 & 139  & 33 \\ 
 2.3 - 2.4  & 13412 & 3900 & 573 & 110 & 5.5 	& 12246 & 3691 & 594 & 125 & 7.7 & 9025 & 2147 & 356 & 115  & 27 \\ 
 2.4 - 2.5  & 12240 & 3452 & 479 & 87& 3.9  	& 10556 & 2969 & 420 & 78 & 3.8 & 8040 & 1835 & 298 & 96  & 22 \\ 
\hline
\\[-1ex]
 ~ & ~  & ~ & ~ & ~ & ~  & $N$ & ~ & ~ & ~  & ~ & ~ & ~  & ~ & ~ & ~\\[1ex]
 ~ & Model 1  & ~ & ~ & ~ & ~ & Model 2  & ~ & ~ & ~ & ~ & Model 3  & ~ & ~ & ~ & ~ \\[1ex]
 0.7 - 1.5 & 30305 & 14063 & 4962 & 2219 & 601  & 27371 & 13773 & 5588 & 2837 & 952 & 19698 & 8389 & 2523 & 1033 & 284  \\
 1.5 - 2.0 & 12188 & 4378 & 958 & 269      & 28     & 12369 & 5095 & 1465 & 537 & 97 & 8393 & 2663 & 551 & 189  & 46  \\
 0.7 - 2.0 & 42493 & 18441 & 5920 & 2488 & 629  & 39740 & 18868 & 7053 & 3374 & 1049 & 28091 & 11052 & 3074 & 1222  & 330 \\
 0.9 - 1.8 & 30708 & 13034 & 3939 & 1527 & 317  & 28225 & 13266 & 4819 & 2216 & 621 & 19995 & 7733 & 2041 & 779  & 203  \\
 0.4 - 1.8 & 48053 & 22775 & 8596 & 4244 & 1489 & 45208 & 23027 & 9699 & 5183 & 2002 & - & - & - & -  & - \\
\hline\hline
\end{tabular}
}
\end{table*}

The redshift distributions ($dN/dz$) at various \Ha\ flux limits ($0.5, 1, 2, 3\times 10^{-16}$ erg cm$^{-2}$ s$^{-1}$) relevant to the {\slshape Euclid} and {\slshape WFIRST-AFTA} surveys are shown in Figure~\ref{fig:hafit}. 
Besides the observed WISP cumulative counts (from their Table 2, interpolating at the \Ha\ flux limits corrected for \nii\ contamination), we show also $dN/dz$ derived using single luminosity functions observed at different redshifts (integrated over the observed redshift range and plotted at the central redshift of each survey).
It is evident that the current scatter in the observed luminosity function at $z>1$, introduces a large uncertainty in the predictions, in particular at bright fluxes.
The differences between our three models are due to the different evolution and parametrization assumed for the luminosity functions. 
In particular, as discussed in previous section, with Model 2 having a brighter $L_\star$ at high redshifts, the predicted $dN/dz$ is higher at bright fluxes at $z>1.2$. 
Clearly, this is a regime where large areas data are almost not available, and {\slshape Euclid} will cover this gap.
At faint fluxes, instead Model 1 and 2 are more similar, sampling the low 
luminosity end of their similar LFs, with Model 1 having a slightly steeper LF and higher $\phi_\star$. 
Model 3 predicts a density of emitters that is a factor from 1.5 to 2.5 lower than the other models from the faint to the bright fluxes considered, at all redshifts.

The new analysis by \citet{Mehta15} of the WISP survey is also shown in Figure \ref{fig:hafit}.
The number densities at $z\sim 2$ have been derived using the [OIII] line luminosity function, and assuming that 
the relation between \Ha\ and [OIII] luminosity does not change significantly over the redshift range.
The expectations are quite high but consistent within the error-bars  with our highest model.

In Table \ref{tab:dNdz} we list the predicted redshift distributions in redshift bins of width of $\Delta z=0.1$, at various limiting fluxes, for the three different models\footnote{Complete tables for the 3 models at limiting fluxes from $0.1$ to $100 \times 10^{-16}$ erg cm$^{-2}$ s$^{-1}$ are available at http://www.bo.astro.it/$\sim$pozzetti/Halpha/Halpha.html}.
In addition we list also the expected numbers for the 3 models at different flux limits and in the typical redshift ranges for future NIR space missions. In particular for the original {\slshape Euclid} wide/deep 
surveys, designed to cover in \Ha\ the redshift range $0.7<z<2.0$ at flux limit about $3/0.5 \times 10^{-16}$ erg cm$^{-2}$ s$^{-1}$, using two grisms (blue+red), we expect about 1200/28000 -- 3400/40000 objects for deg$^2$.
We note that for the wide survey similar number densities can be reached using the same exposure time but a single grism, for example covering the redshift range $0.9<z<1.8$ to a flux limit of $2 \times 10^{-16}$ erg cm$^{-2}$ s$^{-1}$, we expect about 2000-4800 \Ha\ emitters/deg$^2$, therefore in total 30-72 million of sources will be mapped by {\slshape Euclid}.
For the {\slshape Euclid} deep survey an extension of the grism to bluer wavelenghts, i.e. to lower redshift, for example $0.4<z<1.8$, will increase the number densities to about 32000--48000 deg$^{-2}$ and therefore 1.3--2 million of emitters mapped in 40 deg$^2$.

We remind the reader that these predictions are in terms of observed \Ha\ flux, i.e. include intrinsic dust 
extinction in the $H\alpha$ emitters, and is corrected for \nii\ contamination.
However, at the nominal resolution of both {\slshape Euclid} and {\slshape WFIRST-AFTA}, 
these lines will be partially blended; thus,
here we may be underestimating the final detection significance of the galaxies, i.e.  
the sensitivity to \Ha\ flux is better than the single line sensitivity described here, even if not by the 
full factor of the \Ha:(\Ha+\nii) ratio.
For this reason, we expect that our analysis is somewhat conservative.

Finally, future NIR space mission will use slitless spectroscopy and therefore suffer from some degree of 
contamination in the spectra (depending on the rotation angles used), as well as misidentification of 
different emission lines, 
which will decrease the effective numbers of emitters available for science.
We note, 
however, that unlike the WISP survey, both Euclid and WFIRST will use 
multiple dispersion angles to break the degeneracy in which lines from 
different sources at different wavelengths can fall on the same pixel.
Therefore the number of objects computed from the \Ha LF should be reduced by the 
completeness factor before being used in cosmological forecasts.
Preliminary estimates of this factor have been included in the {\slshape Euclid} \citep{2011arXiv1110.3193L} and {\slshape WFIRST-AFTA} \citep{Spergel15}
forecasts, and the estimates of completeness will continue to be refined as the instrument, pipeline, and
simulations are developed. The problem of sample contamination depends on the abundance of other line emitters (i.e. not \Ha), and the data available to reject them -- photometric redshifts and secondary lines. The rejection logic is specific to each survey as it depends on wavelength range, grism resolution (e.g. both {\slshape Euclid} and {\slshape WFIRST-AFTA} would separate the \oiii\ doublet), and which deep imaging filters are available. \citet{2015arXiv150705092P} present an example for {\slshape WFIRST-AFTA} (combined with LSST photometry) that should achieve a low contamination rate, albeit under idealized assumptions. It is presumed that deep spectroscopic training samples will be required to characterize the contamination rate in the cosmology sample.
The redshift completeness factor could also be density dependent, and its estimation will require mock catalogues with clustering. The
models presented here, therefore, will provide a key input to instrument simulations that aim to forecast the completeness of the Euclid and WFIRST-AFTA spectroscopic samples.

\section{Comparison to semi-analytic mock catalogs}
\label{sec:mocks}

We compare our empirical \Ha\ models to \Ha\ number counts
and redshift distributions from mock galaxy catalogues built with the
semi-analytic galaxy formation model GALFORM (e.g. Cole et al. 2000; 
Bower et al. 2006).
The dark matter halo merger trees with which 
GALFORM builds a galaxy catalogue are extracted from two flat 
$\Lambda$CDM simulations of 500Mpc/h aside, differing only by their 
cosmology: 
(i) the Millennium simulation \citep{2005Natur.435..629S} 
with $\Omega_{\rm m}$=0.25, $h$ = 0.73 and $\sigma_8$ = 0.90;
(ii) the MR7 simulation \citep{Guo13} 
with $\Omega_{\rm m}$=0.272, $h$ = 0.704 and $\sigma_8$ = 0.81.
\citet{2013MNRAS.429..556M} 
provides a method for constructing 
lightcone galaxy catalogues from the GALFORM populated 
simulation snapshots, onto which observational selections 
can be applied, like an apparent H-band magnitude limit.
These lightcones come with an extensive list of galaxy 
properties, including the observed and cosmological redshifts, the 
observed magnitudes and rest-frame absolute magnitudes in several
bands, and the observed fluxes and rest-frame luminosities of 
several emission lines.
For the present work, we have analysed lightcones built with the 
\citet{2012MNRAS.426.2142L} GALFORM model\footnote{\citet{2012MNRAS.426.2142L} Euclid lightcones are available from http://community.dur.ac.uk/a.i.merson/lightcones.html}
 (L12 mocks, hereafter) 
and with the \citet{Gonzalez-Perez14} GALFORM model\footnote{\citet{Gonzalez-Perez14} lightcones are available from the Millennium Database, accessible from http://www.icc.dur.ac.uk/data/}
(GP14 mocks hereafter), using respectively the Millennium and MR7 
simulations.
It is essential to point out that the model parameters for 
\citet{2012MNRAS.426.2142L} and \citet{Gonzalez-Perez14} are calibrated 
using mostly local datasets, such as the optical and near-IR galaxy
luminosity functions. In particular, no observational constraints from 
emission line galaxies are used in the calibration process.

We are most interested in the $H$ band magnitude and the H$\alpha$ flux. 
To assign galaxy properties, the stellar population synthesis models from \citet{2003MNRAS.344.1000B} (GISSEL 99 version) 
is used with the \citet{1983ApJ...272...54K} initial mass function 
over the range $0.15 M_\odot < m < 120 M_\odot$. 
To calculate the H$\alpha$ flux, the number of Lyman
continuum photons is computed from the star formation history predicted for
the galaxy. The \citet{1990A&AS...83..501S} models is used to obtain
the line luminosity from the number of continuum photons (T=45000 K, n=10 and Ns=1), testing that the choice of the HII region properties from the Stasinska models
does not have an impact on the number of \Ha\ emitters. 
The dust extinction law is the one for the Milky Way by \citet{Ferrara99}.
Broad-band magnitudes are reported on the AB scale.

In this work, we have analysed the light-cones constructed to emulate the {\slshape Euclid} surveys. 
In particular, in  order to explore the effect of the selection on the density of
H$\alpha$ emitters, we have explored deep mocks selected in magnitude
provided on different areas (100 deg$^2$ limited to $H<27$ for L12 mocks and
20 deg$^2$ limited to $H<27$ or $F_{{\rm H}\alpha}>3\times 10^{-18}$
erg cm$^{-2}$ for GP14 one). 

In Figure~\ref{fig:hacounts} we show the cumulative number densities derived using different mock 
catalogues in the redshift range $0.7<z<1.5$.
The two light-cones, irrespective of the GALFORM version used, underpredict the
cumulative counts at all \Ha\ fluxes explored (from $>10^{-15}$  up to $>10^{-17}$ erg cm$^{-2}$ s$^{-1}$), 
i.e. they are therefore 
in disagreement with the observed counts from WISP survey and with most of the 
counts derived from empirical LF, with the only exception of HiZELS \citet{2013MNRAS.428.1128S}.
We have also tested the effect of limiting magnitudes on the \Ha\ counts from the GP14 mock, finding that 
a mock selected to $H<24$ 
underestimates the density of H$\alpha$ emitters but only at very faint fluxes
($3\times10^{-17}$ erg cm$^{-2}$ s$^{-1}$).

Finally, we compare the mocks with our empirical models.
We find that the mocks predict counts, in the redshift $0.7<z<1.5$, lower than our models; for example the GP14 mock is lower than Model 1 by a factor  2 to 4.5 from faint to bright flux limits.
The L12 mock predictions are even slightly lower than the GP14 mock.

We have also explored the effect of dust extinction, showing the cumulative counts for intrinsic \Ha\ fluxes (i.e. before dust extinction applied) for the L12 mock in Figure ~\ref{fig:hacounts}. In this case the simulated mock predicts very high number densities at bright fluxes ($>10^{-15}$ erg cm$^{-2}$ s$^{-1}$), above all the data available, but agree with the two faintest data from the WISP survey, which are close to the deep {\slshape Euclid} flux limit.
We note that the effect of dust extinction is flux dependent in the L12 mock. However, also
predictions using intrinsic \Ha\ fluxes and applying a dust extinction of 1 magnitude (0.4 dex), as usually applied
reversally in the data, provide counts even lower than mocks shown and flatter than our 3 models and explored data.

In Figure~\ref{fig:hafit} we further analyse the predictions for the redshift distribution,
 at various flux limits, from the light-cones.
It is evident that redshift distribution from mocks, irrespective of the explored flux limits, 
are  consistent with data at low redshift, while they are systematically lower than data at $z>1$, 
despite the large dispersion in the data. The number densities from GP14 mock are lower than our models and 
data by a factor up to 10 at faint fluxes and $z>1.5$.
The L12 mock predictions are even lower than the GP14 ones, in particular at $z>1.5$, where
at all fluxes except for faint ones, the number densities continue to decrease with $z$ and not present
a flattening as in the GP14 mock.
At low redshift, instead,
the mocks cover relatively well the range of number densities predicted by our models, 
being more similar to Model 1, 2 at $z<0.7$.

In addition, we have explored the effect of a brighter H-band magnitude limit ($H<24$ compared to the original $H<27$) in the GP14 mock, finding that it does not affect 
strongly the redshift distribution at all flux explored, but fainter one at high redshift ($z>2$).
Finally, we note that only 
using intrinsic \Ha\ fluxes, i.e. before dust extinction has been applied, 
the L12 mock
predicts a tail in the redshift distribution at high redshift ($z>1.5-2$) and bright flux limits
consistent with our empirical models or even higher at $z>2$
(note however that empirical models and data are not corrected for dust extinction).

We remind the reader that 
the SAMs used here are not calibrated using emission line datasets.
The predictions of mocks for the \HaLF s  have been analysed by \citet{Lagos14} (see their Figure 1).
In a future work (Shi et al., in preparation), we will analyse in detail 
an optimization of the mocks to reproduce empirical \HaLF s,
taking into account also the contribution of AGN, which might affect this comparison.

\begin{figure}
\includegraphics[width=88mm]{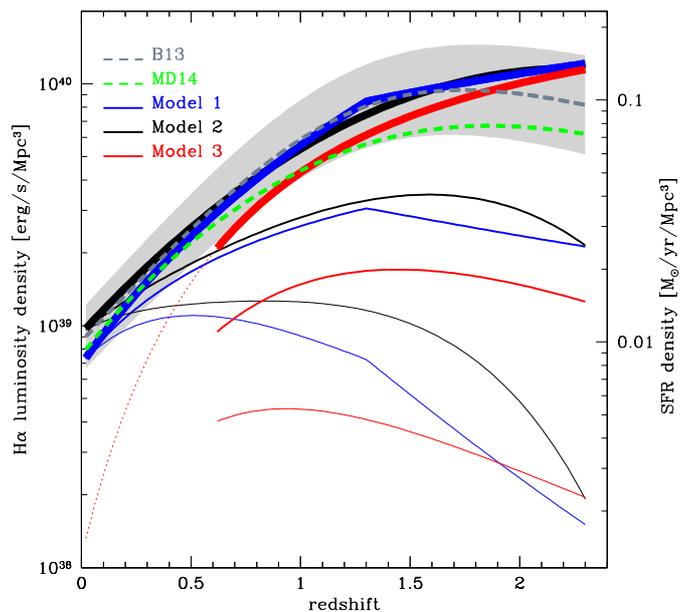}
\caption{\label{fig:sfr1} \Ha\ luminosity density of the Universe as a function of redshift. The solid thick lines show the total luminosity density, whereas the thin solid curves show the luminosity density for emitters at $F>10^{-16}$ erg cm$^{-2}$ s$^{-1}$ (upper set of curves) and $F>3\times 10^{-16}$ erg cm$^{-2}$ s$^{-1}$ (lower set of curves). The different colours are codes for each model (blue, Model 1; black, Model 2; and red, Model 3). Also shown is the calculation of total \Ha\ luminosity density based on the star formation histories of \citet[green dashed]{2014ARA&A..52..415M} and \citet[gray dashed and shaded area]{2013ApJ...770...57B}. On the right axis we report also the SFR density scale for a Chabrier IMF.
}
\end{figure}

\section{\Ha\ luminosity density and star formation history}
\label{sec:sfh}

Finally, we consider the implications of our empirical models for the global \Ha\ luminosity density of the Universe, and the closely related cosmic star formation history.

The \Ha\ luminosity density is shown in Figure~\ref{fig:sfr1}, as predicted by each model. We show the total (integrating the functional form over all luminosities), along with the predictions by each model 
imposing flux limits at $F>10^{-16}$ erg cm$^{-2}$ s$^{-1}$ and $F>3\times 10^{-16}$ erg cm$^{-2}$ s$^{-1}$. 
Note the excellent agreement for the integrated luminosity functions, even though the bright end is quite different for the 3 models (being lower for Model 3). As one can see from the dashed curves, the depth probed by BAO surveys picks up only a portion of the overall \Ha\ emission in the Universe: for example, Models 1, 2, and 3 predict that 31, 39, and 23 per cent respectively of the \Ha\ emission passes the flux cut  $F> 10^{-16}$ erg cm$^{-2}$ s$^{-1}$ at $z=1.5$. At $z=1.5$, to represent half of the overall \Ha\ emission, we would need to lower the flux cut to $(3.1-6.6)\times 10^{-17}$ erg cm$^{-2}$ s$^{-1}$.

Also shown in Figure~\ref{fig:sfr1} is the  observed (i.e. no corrected for extinction) \Ha\ luminosity density
derived from the star formation history  by \citet{2014ARA&A..52..415M} and 
\citet{2013ApJ...770...57B} along with its dispersion. 
We, respectively, use the conversion of $L_{{\rm H}\alpha}/{\rm SFR} = 7.9\times 10^{-42}$ erg s$^{-1}$ $M_\odot^{-1}$ yr \citep{1998ARA&A..36..189K}, 
appropriate for a \citet{1955ApJ...121..161S} initial mass function used by \citet{2014ARA&A..52..415M}, and
adding a factor of 1.7 boost, appropriate for the \citet{2003PASP..115..763C} initial mass function, used by \citet{2013ApJ...770...57B}.
For consistency, the same receipt used to correct for dust extinction in \Ha\ surveys to derive the above SFHs, has been used to correct them back, i.e. the derived \Ha\ luminosity density  has been reduced by a factor of $10^{0.4}$ in 
accordance with the commonly-assumed 1 magnitude of extinction (Hopkins et al. 2004).
This procedure is not an independent check, since  \citet{2013ApJ...770...57B} refer to some of the same data used in this 
paper, 
but does provide an assessment of the overall consistency of the literature, particularly given that \citet{2014ARA&A..52..415M}  and \citet{2013ApJ...770...57B}  consider many other tracers of star formation (i.e. not just \Ha) as well.
 The agreement is within a factor of 2 difference at $z\sim 2$ for one of the star formation histories, and better for other cases -- we consider this good, given the uncertainties in the extrapolation to fluxes lower than that covered by \Ha\ surveys. 
We further note that the 
agreement is still recovered if we consider a more sophisticated treatment of the dust extinction, 
varying it with redshift as derived from the ratio between FUV and FIR luminosity densities \citep{Burgarella13}. 
However, this procedure introduces additional and 
uncertain assumptions on the dust extinction law and on the ratio between the extinction in the continuum and in 
the emission lines \citep{Calzetti00}.

Finally, we note that on the contrary the SAMs considered in this paper predict a star formation density 
below the values deduced from the observations at $0.3<z<2$ (see \citet{Lagos14}, Figure 3). We emphasize here 
that the observed/exctincted \Ha\ luminosity density is inferred after applying a correction for dust extinction 
and after extrapolation down to faint unobserved \Ha\ luminosities, introducing therefore further 
uncertainties in the comparison with models.

\section{Summary}
\label{sec:discussion}

The H$\alpha$ luminosity function is a key ingredient for forecasts for future dark energy surveys, especially at $z\gtrsim 1$ where blind emission-line selection is one of the most efficient ways to build large statistical samples of galaxies with known redshifts. We have collected the main observational results from the literature and provided three empirical H$\alpha$ luminosity function models. Models 1 and 2 have the advantage of combining the largest amount of data over the widest redshift range, whereas Model 3 focuses only on fitting 
the range of redshift and flux most relevant to {\slshape Euclid} and {\slshape WFIRST-AFTA}, but covered by more sparse and uncertain data.

The three model H$\alpha$ luminosity functions are qualitatively similar, but there are differences of up to a factor of 3 (ratio of highest to lowest) in the most discrepant parts of Table~\ref{tab:dNdz}. This is despite the small aggregate statistical errors (for example $\pm$17 per cent at $2\sigma$ for Model 3 in the redshift range of $0.9<z<1.8$ 
and $F_{{\rm H}\alpha}>2\times 10^{-16}$ erg cm$^{-2}$ s$^{-1}$). Some of this is due to real differences in the input data sets. In particular, our investigations of the input data in Model 3 show that minor details in the fits (such as the treatment of asymmetric error bars and the finite width of luminosity and redshift bins) as well as cosmic variance affect the outcome by more than the statistical errors in the fits. All of these models predict significantly fewer \Ha\ emitters than were anticipated several years ago. However, even according to our most conservative model, the upcoming space missions {\slshape Euclid} and {\slshape WFIRST-AFTA} will chart the three-dimensional positions of tens of millions of galaxies at $z\gtrsim 0.9$, a spectacular advance over the capabilities of present-day redshift surveys. 
For instance, covering the redshift range $0.9<z<1.8$ to a flux limit of $2 \times 10^{-16}$ erg cm$^{-2}$ s$^{-1}$, we expect about 2000-4800 \Ha\ emitters/deg$^2$, therefore in total 30-72 million of sources will be mapped over 15 000 deg$^2$ by the {\slshape Euclid} wide survey and 1.3-2 million of emitters will be mapped in 40 deg$^2$ by the {\slshape Euclid} deep survey in the range $0.4<z<1.8$ at fluxes above $0.5 \times 10^{-16}$ erg cm$^{-2}$ s$^{-1}$.
At the {\slshape WFIRST-AFTA} sensitivity, we predict in the redshift range $1<z<1.9$ about 16 to 26 million of galaxies at fluxes above  $\sim 1 \times 10^{-16}$ erg cm$^{-2}$ s$^{-1}$ over 2200 deg$^2$.
The models presented here also provide a key input for the scientific optimization of the survey parameters of these missions and for cosmological forecasts  from the spectroscopic samples of {\slshape Euclid} and {\slshape WFIRST-AFTA}.
The \Ha LFs derived here must be folded through instrument performance, observing strategy and completeness, and modelling of the galaxy power spectrum in order to arrive at predicted BAO constraints. The previous {\slshape Euclid} forecasts 
\citep{2013LRR....16....6A} are currently being updated with the new \Ha LFs and updated instrument parameters, and we anticipate that the public documents will be updated soon. The \Ha LFs presented here have already been incorporated
in the most recent {\slshape WFIRST-AFTA} science report \citep{Spergel15}.

\begin{acknowledgements}

We wish to thank James Colbert, David Sobral, Lin Yan, Gunawardhana Madusha and Ly Chun for their help with the input luminosity function data for this paper. 
We thank Bianca Garilli, Gigi Guzzo, Yun Wang, Will Percival, Claudia Scarlata and Gianni Zamorani to stimulate this work and for useful discussions and comments. 
L.P.  and A.C. acknowledge financial contributions by grants ASI/INAF
I/023/12/0 and PRIN MIUR 2010-2011 ``The dark Universe and the cosmic
evolution of baryons: from current surveys to Euclid.''
C.H. has been supported by the United States Department of Energy under contract DE-FG03-02-ER40701, the David and Lucile Packard Foundation, the Simons Foundation, and the Alfred P. Sloan Foundation.
This work was supported by the STFC through grant number  ST/K003305/1.
J.E.G. thanks the Royal Society.
P.N. acknowledges the support of the Royal Society through the award
of a University Research Fellowship and the European Research
Council, through receipt of a Starting Grant (DEGAS-259586).
C.M.B., P.N. and S.D. acknowledge the support of the Science and Technology
Facilities Council (ST/L00075X/1).
The simulations results used the DiRAC Data Centric system at Durham University 
and funded by BIS National E-infrastructure capital grant ST/K00042X/1, 
STFC capital grants ST/H008519/1 and ST/K00087X/1, STFC DiRAC Operations 
grant ST/K003267/1 and Durham University. 

\end{acknowledgements}


\bibliographystyle{aa} 

\begin{thebibliography}{56}
\expandafter\ifx\csname natexlab\endcsname\relax\def\natexlab#1{#1}\fi

\bibitem[{{Albrecht} {et~al.}(2006){Albrecht}, {Bernstein}, {Cahn}, {Freedman},
  {Hewitt}, {Hu}, {Huth}, {Kamionkowski}, {Kolb}, {Knox}, {Mather}, {Staggs},
  \& {Suntzeff}}]{2006astro.ph..9591A}
{Albrecht}, A., {Bernstein}, G., {Cahn}, R., {et~al.} 2006, arXiv:astro-ph/0609591

\bibitem[{{Amendola} {et~al.}(2013)}]{2013LRR....16....6A}
{Amendola}, L. {et~al.} 2013, {\it Living Reviews in Relativity},
16, 6

\bibitem[{{Anderson} {et~al.}(2014){Anderson}, {Aubourg}, {Bailey}, {Beutler},
  {Bhardwaj}, {Blanton}, {Bolton}, {Brinkmann}, {Brownstein}, {Burden},
  {Chuang}, {Cuesta}, {Dawson}, {Eisenstein}, {Escoffier}, {Gunn}, {Guo}, {Ho},
  {Honscheid}, {Howlett}, {Kirkby}, {Lupton}, {Manera}, {Maraston}, {McBride},
  {Mena}, {Montesano}, {Nichol}, {Nuza}, {Olmstead}, {Padmanabhan},
  {Palanque-Delabrouille}, {Parejko}, {Percival}, {Petitjean}, {Prada},
  {Price-Whelan}, {Reid}, {Roe}, {Ross}, {Ross}, {Sabiu}, {Saito}, {Samushia},
  {S{\'a}nchez}, {Schlegel}, {Schneider}, {Scoccola}, {Seo}, {Skibba},
  {Strauss}, {Swanson}, {Thomas}, {Tinker}, {Tojeiro}, {Maga{\~n}a}, {Verde},
  {Wake}, {Weaver}, {Weinberg}, {White}, {Xu}, {Y{\`e}che}, {Zehavi}, \&
  {Zhao}}]{2014MNRAS.441...24A}
{Anderson}, L., {Aubourg}, {\'E}., {Bailey}, S., {et~al.} 2014, \mnras, 441, 24

\bibitem[{{Behroozi} {et~al.}(2013){Behroozi}, {Wechsler}, \&
  {Conroy}}]{2013ApJ...770...57B}
{Behroozi}, P.~S., {Wechsler}, R.~H., \& {Conroy}, C. 2013, \apj, 770, 57

\bibitem[{{Blake} {et~al.}(2011{\natexlab{a}}){Blake}, {Davis}, {Poole},
  {Parkinson}, {Brough}, {Colless}, {Contreras}, {Couch}, {Croom},
  {Drinkwater}, {Forster}, {Gilbank}, {Gladders}, {Glazebrook}, {Jelliffe},
  {Jurek}, {Li}, {Madore}, {Martin}, {Pimbblet}, {Pracy}, {Sharp}, {Wisnioski},
  {Woods}, {Wyder}, \& {Yee}}]{2011MNRAS.415.2892B}
{Blake}, C., {Davis}, T., {Poole}, G.~B., {et~al.} 2011{\natexlab{a}}, \mnras,
  415, 2892

\bibitem[{{Blake} {et~al.}(2011{\natexlab{b}}){Blake}, {Kazin}, {Beutler},
  {Davis}, {Parkinson}, {Brough}, {Colless}, {Contreras}, {Couch}, {Croom},
  {Croton}, {Drinkwater}, {Forster}, {Gilbank}, {Gladders}, {Glazebrook},
  {Jelliffe}, {Jurek}, {Li}, {Madore}, {Martin}, {Pimbblet}, {Poole}, {Pracy},
  {Sharp}, {Wisnioski}, {Woods}, {Wyder}, \& {Yee}}]{2011MNRAS.418.1707B}
{Blake}, C., {Kazin}, E.~A., {Beutler}, F., {et~al.} 2011{\natexlab{b}},
  \mnras, 418, 1707

\bibitem[{{Bond} {et~al.}(1998){Bond}, {Jaffe}, \&
  {Knox}}]{1998PhRvD..57.2117B}
{Bond}, J.~R., {Jaffe}, A.~H., \& {Knox}, L. 1998, \prd, 57, 2117

\bibitem[{{Bond} {et~al.}(2000){Bond}, {Jaffe}, \&
  {Knox}}]{2000ApJ...533...19B}
{Bond}, J.~R., {Jaffe}, A.~H., \& {Knox}, L. 2000, \apj, 533, 19

\bibitem[{{Bruzual} \& {Charlot}(2003)}]{2003MNRAS.344.1000B}
{Bruzual}, G. \& {Charlot}, S. 2003, \mnras, 344, 1000

\bibitem[{{Burgarella} {et~al.}(2013)}]{Burgarella13}
{Burgarella}, D., {Buat}, V., {Gruppioni}, C., {Cucciati}, O., {et~al.} 2013, \aap, 554, 70

\bibitem[{{Calzetti} {et~al.}(2000)}]{Calzetti00}
{Calzetti}, D., {Armus}, L., {Bohlin}, R. ~C., {Kinney}, A. L., {et~al.} 2000, \apj, 533, 682

\bibitem[{{Chabrier}(2003)}]{2003PASP..115..763C}
{Chabrier}, G. 2003, \pasp, 115, 763

\bibitem[{{Colbert} {et~al.}(2013){Colbert}, {Teplitz}, {Atek}, {Bunker},
  {Rafelski}, {Ross}, {Scarlata}, {Bedregal}, {Dominguez}, {Dressler}, {Henry},
  {Malkan}, {Martin}, {Masters}, {McCarthy}, \& {Siana}}]{2013ApJ...779...34C}
{Colbert}, J.~W., {Teplitz}, H., {Atek}, H., {et~al.} 2013, \apj, 779, 34

\bibitem[{{Cole} {et~al.}(2005){Cole}, {Percival}, {Peacock}, {Norberg},
  {Baugh}, {Frenk}, {Baldry}, {Bland-Hawthorn}, {Bridges}, {Cannon}, {Colless},
  {Collins}, {Couch}, {Cross}, {Dalton}, {Eke}, {De Propris}, {Driver},
  {Efstathiou}, {Ellis}, {Glazebrook}, {Jackson}, {Jenkins}, {Lahav}, {Lewis},
  {Lumsden}, {Maddox}, {Madgwick}, {Peterson}, {Sutherland}, \&
  {Taylor}}]{2005MNRAS.362..505C}
{Cole}, S., {Percival}, W.~J., {Peacock}, J.~A., {et~al.} 2005, \mnras, 362,
  505

\bibitem[{{Eisenstein} {et~al.}(2005){Eisenstein}, {Zehavi}, {Hogg},
  {Scoccimarro}, {Blanton}, {Nichol}, {Scranton}, {Seo}, {Tegmark}, {Zheng},
  {Anderson}, {Annis}, {Bahcall}, {Brinkmann}, {Burles}, {Castander},
  {Connolly}, {Csabai}, {Doi}, {Fukugita}, {Frieman}, {Glazebrook}, {Gunn},
  {Hendry}, {Hennessy}, {Ivezi{\'c}}, {Kent}, {Knapp}, {Lin}, {Loh}, {Lupton},
  {Margon}, {McKay}, {Meiksin}, {Munn}, {Pope}, {Richmond}, {Schlegel},
  {Schneider}, {Shimasaku}, {Stoughton}, {Strauss}, {SubbaRao}, {Szalay},
  {Szapudi}, {Tucker}, {Yanny}, \& {York}}]{2005ApJ...633..560E}
{Eisenstein}, D.~J., {Zehavi}, I., {Hogg}, D.~W., {et~al.} 2005, \apj, 633, 560

\bibitem[{{Ferrara} {et~al.}(1999){Ferrara}, {Bianchi}, {Cimatti}, \& {Giovanardi}}]{Ferrara99}
{Ferrara}, A., {Bianchi}, S., {Cimatti}, A., \& {Giovanardi}, C. 1999, \apjs, 123, 437

\bibitem[{{Gallego} {et~al.}(1995){Gallego}, {Zamorano}, {Aragon-Salamanca}, \&
  {Rego}}]{1995ApJ...455L...1G}
{Gallego}, J., {Zamorano}, J., {Aragon-Salamanca}, A., \& {Rego}, M. 1995,
  \apjl, 455, L1

\bibitem[{{Geach} {et~al.}(2010){Geach}, {Cimatti}, {Percival}, {Wang},
  {Guzzo}, {Zamorani}, {Rosati}, {Pozzetti}, {Orsi}, {Baugh}, {Lacey},
  {Garilli}, {Franzetti}, {Walsh}, \& {K{\"u}mmel}}]{2010MNRAS.402.1330G}
{Geach}, J.~E., {Cimatti}, A., {Percival}, W., {et~al.} 2010, \mnras, 402, 1330

\bibitem[{{Geach} {et~al.}(2008){Geach}, {Smail}, {Best}, {Kurk}, {Casali},
  {Ivison}, \& {Coppin}}]{2008MNRAS.388.1473G}
{Geach}, J.~E., {Smail}, I., {Best}, P.~N., {et~al.} 2008, \mnras, 388, 1473

\bibitem[{{Geach} {et~al.}(2012){Geach}, {Sobral}, {Hickox}, {Wake}, {Smail},
  {Best}, {Baugh}, \& {Stott}}]{2012MNRAS.426..679G}
{Geach}, J.~E., {Sobral}, D., {Hickox}, R.~C., {et~al.} 2012, \mnras, 426, 679

\bibitem[{{Green} {et~al.}(2011){Green}, {Schechter}, {Baltay}, {Bean},
  {Bennett}, {Brown}, {Conselice}, {Donahue}, {Gaudi}, {Lauer}, {Perlmutter},
  {Rauscher}, {Rhodes}, {Roellig}, {Stern}, {Sumi}, {Tanner}, {Wang}, {Wright},
  {Gehrels}, {Sambruna}, \& {Traub}}]{Green11}
{Green}, J., {Schechter}, P., {Baltay}, C., {et~al.} 2011, arXiv:1108.1374

\bibitem[{{Green} {et~al.}(2012){Green}, {Schechter}, {Baltay}, {Bean}, \& {etal}}]{Green12}
{Green}, J., {Schechter}, P., {Baltay}, C., {et~al.} 2012, arXiv:1208.4012

\bibitem[{{Gonzalez-Perez} {et~al.}(2014){Gonzalez-Perez}, {Lacey}, {Baugh}, {Lagos}, \& {etal}}]{Gonzalez-Perez14}
{Gonzalez-Perez}, V., {Lacey}, C.~G., {Baugh}, C.~M., {et~al.} 2014, \mnras, 439, 264

\bibitem[{{Gunawardhana} {et~al.}(2013)}]{Gunawardhana13}
{Gunawardhana}, M.L.P., {Hopkins}, A.M., {Bland-Hawthorn}, J., {et al.} 2013, \mnras, 433, 2764

\bibitem[{{Gunawardhana} {et~al.}(2015)}]{Gunawardhana15}
{Gunawardhana}, M.L.P., {Hopkins}, A.M., {Taylor}, E.N., {et al.} 2015, \mnras, 447, 875

\bibitem[{{Guo} {et~al.}(2013)}]{Guo13}
{Guo}, Q., {White}, S., {Angulo}, R., {Henriques}, B., {et al.} 2013, \mnras, 428, 1351

\bibitem[{{Guzzo} {et~al.}(2008)}]{2008Natur.451..541G}
{Guzzo}, L., {Pierleoni}, M., {Meneux}, B., {Branchini}, E., {et al.} 2008, \nat, 451, 541

\bibitem[{{Hayes} {et~al.}(2010){Hayes}, {Schaerer}, \&
  {{\"O}stlin}}]{2010A&A...509L...5H}
{Hayes}, M., {Schaerer}, D., \& {{\"O}stlin}, G. 2010, \aap, 509, L5

\bibitem[{{Hirata} {et~al.}(2012){Hirata}, {Gehrels}, {Kneib}, {Kruk},
  {Rhodes}, {Wang}, \& {Zoubian}}]{2012arXiv1204.5151H}
{Hirata}, C.~M., {Gehrels}, N., {Kneib}, J.-P., {et~al.} 2012, arXiv:1204.5151


\bibitem[{{Hopkins} {et~al.}(2000){Hopkins}, {Connolly}, \&
  {Szalay}}]{2000AJ....120.2843H}
{Hopkins}, A.~M., {Connolly}, A.~J., \& {Szalay}, A.~S. 2000, \aj, 120, 2843

\bibitem[{{Hopkins}(2004){Hopkins}}]{Hopkins2004}
{Hopkins}, A.~M. 2004, \apj, 615, 209

\bibitem[{{Ilbert} {et~al.}(2009){Ilbert}, {Capak}, {Salvato}, {Aussel},
  {McCracken}, {Sanders}, {Scoville}, {Kartaltepe}, {Arnouts}, {Le Floc'h},
  {Mobasher}, {Taniguchi}, {Lamareille}, {Leauthaud}, {Sasaki}, {Thompson},
  {Zamojski}, {Zamorani}, {Bardelli}, {Bolzonella}, {Bongiorno}, {Brusa},
  {Caputi}, {Carollo}, {Contini}, {Cook}, {Coppa}, {Cucciati}, {de la Torre},
  {de Ravel}, {Franzetti}, {Garilli}, {Hasinger}, {Iovino}, {Kampczyk},
  {Kneib}, {Knobel}, {Kovac}, {Le Borgne}, {Le Brun}, {F{\`e}vre}, {Lilly},
  {Looper}, {Maier}, {Mainieri}, {Mellier}, {Mignoli}, {Murayama}, {Pell{\`o}},
  {Peng}, {P{\'e}rez-Montero}, {Renzini}, {Ricciardelli}, {Schiminovich},
  {Scodeggio}, {Shioya}, {Silverman}, {Surace}, {Tanaka}, {Tasca}, {Tresse},
  {Vergani}, \& {Zucca}}]{2009ApJ...690.1236I}
{Ilbert}, O., {Capak}, P., {Salvato}, M., {et~al.} 2009, \apj, 690, 1236

\bibitem[{{Kaiser}(1987)}]{1987MNRAS.227....1K}
{Kaiser}, N. 1987, \mnras, 227, 1

\bibitem[{{Kazin} {et~al.}(2014){Kazin}, {Koda}, {Blake}, {Padmanabhan},
  {Brough}, {Colless}, {Contreras}, {Couch}, {Croom}, {Croton}, {Davis},
  {Drinkwater}, {Forster}, {Gilbank}, {Gladders}, {Glazebrook}, {Jelliffe},
  {Jurek}, {Li}, {Madore}, {Martin}, {Pimbblet}, {Poole}, {Pracy}, {Sharp},
  {Wisnioski}, {Woods}, {Wyder}, \& {Yee}}]{2014MNRAS.441.3524K}
{Kazin}, E.~A., {Koda}, J., {Blake}, C., {et~al.} 2014, \mnras, 441, 3524

\bibitem[{{Kazin} {et~al.}(2013){Kazin}, {S{\'a}nchez}, {Cuesta}, {Beutler},
  {Chuang}, {Eisenstein}, {Manera}, {Padmanabhan}, {Percival}, {Prada}, {Ross},
  {Seo}, {Tinker}, {Tojeiro}, {Xu}, {Brinkmann}, {Joel}, {Nichol}, {Schlegel},
  {Schneider}, \& {Thomas}}]{2013MNRAS.435...64K}
{Kazin}, E.~A., {S{\'a}nchez}, A.~G., {Cuesta}, A.~J., {et~al.} 2013, \mnras,
  435, 64

\bibitem[{{Kennicutt}(1983)}]{1983ApJ...272...54K}
{Kennicutt}, Jr., R.~C. 1983, \apj, 272, 54

\bibitem[{{Kennicutt}(1998)}]{1998ARA&A..36..189K}
{Kennicutt}, Jr., R.~C. 1998, \araa, 36, 189

\bibitem[{{Kennicutt} {et~al.}(2009){Kennicutt}, {Hao}, {Calzetti},
  {Moustakas}, {Dale}, {Bendo}, {Engelbracht}, {Johnson}, \&
  {Lee}}]{2009ApJ...703.1672K}
{Kennicutt}, Jr., R.~C., {Hao}, C.-N., {Calzetti}, D., {et~al.} 2009, \apj,
  703, 1672

\bibitem[{{Lagos} {et~al.}(2012){Lagos}, {Bayet}, {Baugh}, {Lacey}, {Bell},
  {Fanidakis}, \& {Geach}}]{2012MNRAS.426.2142L}
{Lagos}, C.~d.~P., {Bayet}, E., {Baugh}, C.~M., {et~al.} 2012, \mnras, 426,
  2142

\bibitem[{{Lagos} {et~al.}(2014){Lagos}, 
{Baugh}}]{Lagos14}
{Lagos}, C.~d.~P., {Baugh}, C.~M., {Zwaan}, M.~A., {et~al.} 2014, \mnras, 440, 920
  ...

\bibitem[{{Laureijs} {et~al.}(2011){Laureijs}, {Amiaux}, {Arduini},
  {Augu{\`e}res}, {Brinchmann}, {Cole}, {Cropper}, {Dabin}, {Duvet}, {Ealet},
  \& et~al.}]{2011arXiv1110.3193L}
{Laureijs}, R., {Amiaux}, J., {Arduini}, S., {et~al.} 2011, arXiv:1110.3193


\bibitem[{{Lee} {et~al.}(2012){Lee}, {Ly}, {Spitler}, {Labb{\'e}}, {Salim},
  {Persson}, {Ouchi}, {Dale}, {Monson}, \& {Murphy}}]{2012PASP..124..782L}
{Lee}, J.~C., {Ly}, C., {Spitler}, L., {et~al.} 2012, \pasp, 124, 782

\bibitem[{{Ly} {et~al.}(2011){Ly}, {Lee}, {Dale}, {Momcheva}, {Salim},
  {Staudaher}, {Moore}, \& {Finn}}]{2011ApJ...726..109L}
{Ly}, C., {Lee}, J.~C., {Dale}, D.~A., {et~al.} 2011, \apj, 726, 109

\bibitem[{{Ly} {et~al.}(2007){Ly}, {Malkan}, {Kashikawa}, {Shimasaku}, {Doi},
  {Nagao}, {Iye}, {Kodama}, {Morokuma}, \& {Motohara}}]{2007ApJ...657..738L}
{Ly}, C., {Malkan}, M.~A., {Kashikawa}, N., {et~al.} 2007, \apj, 657, 738

\bibitem[{{Madau} \& {Dickinson}(2014)}]{2014ARA&A..52..415M}
{Madau}, P. \& {Dickinson}, M. 2014, \araa, 52, 415

\bibitem[{{Mehta} {et~al.}(2015){Mehta}, {Scarlata}, {Colbert}, \& {etal}}]{Mehta15}
{Mehta}, V., {Scarlata}, C., {Colbert}, J.~W., {et~al.} 2015, \apj, 811, 141

\bibitem[{{Merson} {et~al.}(2013){Merson}, {Baugh}, {Helly}, {Gonzalez-Perez},
  {Cole}, {Bielby}, {Norberg}, {Frenk}, {Benson}, {Bower}, {Lacey}, \&
  {Lagos}}]{2013MNRAS.429..556M}
{Merson}, A.~I., {Baugh}, C.~M., {Helly}, J.~C., {et~al.} 2013, \mnras, 429,
  556

\bibitem[{{Orsi} {et~al.}(2010){Orsi}, {Baugh}, {Lacey}, {Cimatti}, {Wang}, \&
  {Zamorani}}]{2010MNRAS.405.1006O}
{Orsi}, A., {Baugh}, C.~M., {Lacey}, C.~G., {et~al.} 2010, \mnras, 405, 1006

\bibitem[{{Padmanabhan} {et~al.}(2012){Padmanabhan}, {Xu}, {Eisenstein},
  {Scalzo}, {Cuesta}, {Mehta}, \& {Kazin}}]{2012MNRAS.427.2132P}
{Padmanabhan}, N., {Xu}, X., {Eisenstein}, D.~J., {et~al.} 2012, \mnras, 427,
  2132

\bibitem[{{Percival} {et~al.}(2007){Percival}, {Cole}, {Eisenstein}, {Nichol},
  {Peacock}, {Pope}, \& {Szalay}}]{2007MNRAS.381.1053P}
{Percival}, W.~J., {Cole}, S., {Eisenstein}, D.~J., {et~al.} 2007, \mnras, 381,
  1053

\bibitem[{{Percival} {et~al.}(2010){Percival}, {Reid}, {Eisenstein}, {Bahcall},
  {Budavari}, {Frieman}, {Fukugita}, {Gunn}, {Ivezi{\'c}}, {Knapp}, {Kron},
  {Loveday}, {Lupton}, {McKay}, {Meiksin}, {Nichol}, {Pope}, {Schlegel},
  {Schneider}, {Spergel}, {Stoughton}, {Strauss}, {Szalay}, {Tegmark},
  {Vogeley}, {Weinberg}, {York}, \& {Zehavi}}]{2010MNRAS.401.2148P}
{Percival}, W.~J., {Reid}, B.~A., {Eisenstein}, D.~J., {et~al.} 2010, \mnras,
  401, 2148

\bibitem[{{Perlmutter} {et~al.}(1999){Perlmutter}, {Aldering}, {Goldhaber},
  {Knop}, {Nugent}, {Castro}, {Deustua}, {Fabbro}, {Goobar}, {Groom}, {Hook},
  {Kim}, {Kim}, {Lee}, {Nunes}, {Pain}, {Pennypacker}, {Quimby}, {Lidman},
  {Ellis}, {Irwin}, {McMahon}, {Ruiz-Lapuente}, {Walton}, {Schaefer}, {Boyle},
  {Filippenko}, {Matheson}, {Fruchter}, {Panagia}, {Newberg}, {Couch}, \&
  {Project}}]{1999ApJ...517..565P}
{Perlmutter}, S., {Aldering}, G., {Goldhaber}, G., {et~al.} 1999, \apj, 517,
  565

\bibitem[{{Pullen} {et~al.}(2016)}]{2015arXiv150705092P}
{Pullen}, A., {Hirata}, C., {Dor\'e}, O., {Raccanelli}, A. 2016, \pasj, 68, 12


\bibitem[{{Reddy} {et~al.}(2008){Reddy}, {Steidel}, {Pettini}, {Adelberger},
  {Shapley}, {Erb}, \& {Dickinson}}]{2008ApJS..175...48R}
{Reddy}, N.~A., {Steidel}, C.~C., {Pettini}, M., {et~al.} 2008, \apjs, 175, 48

\bibitem[{{Reddy} \& {Steidel}(2009)}]{2009ApJ...692..778R}
{Reddy}, N.~A., {Steidel}, C.~C. 2009, \apj, 692, 778

\bibitem[{{Riess} {et~al.}(1998){Riess}, {Filippenko}, {Challis},
  {Clocchiatti}, {Diercks}, {Garnavich}, {Gilliland}, {Hogan}, {Jha},
  {Kirshner}, {Leibundgut}, {Phillips}, {Reiss}, {Schmidt}, {Schommer},
  {Smith}, {Spyromilio}, {Stubbs}, {Suntzeff}, \&
  {Tonry}}]{1998AJ....116.1009R}
{Riess}, A.~G., {Filippenko}, A.~V., {Challis}, P., {et~al.} 1998, \aj, 116,
  1009

\bibitem[{{Salpeter}(1955)}]{1955ApJ...121..161S}
{Salpeter}, E.~E. 1955, \apj, 121, 161

\bibitem[{{Saunders} {et~al.}(1990){Saunders}, {RR}, \& {etal}}]{Saunders90}
{Saunders}, W., {Rowan-Robinson}, M., {Lawrence} A. 1990, \mnras, 242, 318

\bibitem[{{Schechter}(1976)}]{1976ApJ...203..297S}
{Schechter}, P. 1976, \apj, 203, 297

\bibitem[{{Shim} {et~al.}(2009){Shim}, {Colbert}, {Teplitz}, {Henry}, {Malkan},
  {McCarthy}, \& {Yan}}]{2009ApJ...696..785S}
{Shim}, H., {Colbert}, J., {Teplitz}, H., {et~al.} 2009, \apj, 696, 785

\bibitem[{{Shioya} {et~al.}(2008){Shioya}, {Taniguchi}, {Sasaki}, {Nagao},
  {Murayama}, {Takahashi}, {Ajiki}, {Ideue}, {Mihara}, {Nakajima}, {Scoville},
  {Mobasher}, {Aussel}, {Giavalisco}, {Guzzo}, {Hasinger}, {Impey}, {Le
  F{\`e}vre}, {Lilly}, {Renzini}, {Rich}, {Sanders}, {Schinnerer}, {Shopbell},
  {Leauthaud}, {Kneib}, {Rhodes}, \& {Massey}}]{2008ApJS..175..128S}
{Shioya}, Y., {Taniguchi}, Y., {Sasaki}, S.~S., {et~al.} 2008, \apjs, 175, 128

\bibitem[{{Sobral} {et~al.}(2009){Sobral}, {Best}, {Geach}, {Smail}, {Kurk},
  {Cirasuolo}, {Casali}, {Ivison}, {Coppin}, \& {Dalton}}]{2009MNRAS.398...75S}
{Sobral}, D., {Best}, P.~N., {Geach}, J.~E., {et~al.} 2009, \mnras, 398, 75

\bibitem[{{Sobral} {et~al.}(2012){Sobral}, {Best}, {Matsuda}, {Smail}, {Geach},
  \& {Cirasuolo}}]{2012MNRAS.420.1926S}
{Sobral}, D., {Best}, P.~N., {Matsuda}, Y., {et~al.} 2012, \mnras, 420, 1926

\bibitem[{{Sobral} {et~al.}(2015){Sobral}, {Matthee}, {Best}, \& {etal}}]{Sobral15}
{Sobral}, D., {Matthee}, J., {Best}, P.~N., {et~al.} 2015, \mnras, 451, 2303

\bibitem[{{Sobral} {et~al.}(2013){Sobral}, {Smail}, {Best}, {Geach}, {Matsuda},
  {Stott}, {Cirasuolo}, \& {Kurk}}]{2013MNRAS.428.1128S}
{Sobral}, D., {Smail}, I., {Best}, P.~N., {et~al.} 2013, \mnras, 428, 1128

\bibitem[{{Spergel} {et~al.}(2015){Springel}, {Gehrels}, {Baltay},
  \& {etal}}]{Spergel15}
{Spergel}, D., {Gehrels}, N., {Baltay}, D., {et~al.} 2015, arXiv:astro-ph/1503.03757

\bibitem[{{Springel} {et~al.}(2005){Springel}, {White}, {Jenkins}, {Frenk},
  {Yoshida}, {Gao}, {Navarro}, {Thacker}, {Croton}, {Helly}, {Peacock}, {Cole},
  {Thomas}, {Couchman}, {Evrard}, {Colberg}, \& {Pearce}}]{2005Natur.435..629S}
{Springel}, V., {White}, S.~D.~M., {Jenkins}, A., {et~al.} 2005, \nat, 435, 629

\bibitem[{{Stasi{\'n}ska}(1990)}]{1990A&AS...83..501S}
{Stasi{\'n}ska}, G. 1990, \aaps, 83, 501

\bibitem[{{Steidel} {et~al.}(2014)}]{2014ApJ...795..165S}
{Steidel}, C. et~al. 2014, \apj, 795, 165

\bibitem[{{Storey} \& {Zeippen}(2000)}]{2000MNRAS.312..813S}
{Storey}, P. \& {Zeippen}, C.~J., 2000, \mnras, 312, 813

\bibitem[{{Takahashi} {et~al.}(2007){Takahashi}, {Shioya}, {Taniguchi},
  {Murayama}, {Ajiki}, {Sasaki}, {Koizumi}, {Nagao}, {Scoville}, {Mobasher},
  {Aussel}, {Capak}, {Carilli}, {Ellis}, {Garilli}, {Giavalisco}, {Guzzo},
  {Hasinger}, {Impey}, {Kitzbichler}, {Koekemoer}, {Le F{\`e}vre}, {Lilly},
  {Maccagni}, {Renzini}, {Rich}, {Sanders}, {Schinnerer}, {Scodeggio},
  {Shopbell}, {Smol{\v c}i{\'c}}, {Tribiano}, {Ideue}, \&
  {Mihara}}]{2007ApJS..172..456T}
{Takahashi}, M.~I., {Shioya}, Y., {Taniguchi}, Y., {et~al.} 2007, \apjs, 172,
  456

\bibitem[{{Tresse} \& {Maddox}(1998)}]{1998ApJ...495..691T}
{Tresse}, L. \& {Maddox}, S.~J. 1998, \apj, 495, 691

\bibitem[{{Tresse} {et~al.}(2002){Tresse}, {Maddox}, {Le F{\`e}vre}, \&
  {Cuby}}]{2002MNRAS.337..369T}
{Tresse}, L., {Maddox}, S.~J., {Le F{\`e}vre}, O., \& {Cuby}, J.-G. 2002,
  \mnras, 337, 369

\bibitem[{{Verde} {et~al.}(2003){Verde}, {Peiris}, {Spergel}, {Nolta},
  {Bennett}, {Halpern}, {Hinshaw}, {Jarosik}, {Kogut}, {Limon}, {Meyer},
  {Page}, {Tucker}, {Wollack}, \& {Wright}}]{2003ApJS..148..195V}
{Verde}, L., {Peiris}, H.~V., {Spergel}, D.~N., {et~al.} 2003, \apjs, 148, 195

\bibitem[{{Xu} {et~al.}(2012){Xu}, {Padmanabhan}, {Eisenstein}, {Mehta}, \&
  {Cuesta}}]{2012MNRAS.427.2146X}
{Xu}, X., {Padmanabhan}, N., {Eisenstein}, D.~J., {Mehta}, K.~T., \& {Cuesta},
  A.~J. 2012, \mnras, 427, 2146

\bibitem[{{Yan} {et~al.}(1999){Yan}, {McCarthy}, {Freudling}, {Teplitz},
  {Malumuth}, {Weymann}, \& {Malkan}}]{1999ApJ...519L..47Y}
{Yan}, L., {McCarthy}, P.~J., {Freudling}, W., {et~al.} 1999, \apjl, 519, L47

\end{thebibliography}

\begin{appendix}

\section{Cosmic variance}
\label{app:cv}

This appendix describes the treatment of cosmic variance in the Model 3 fits.

In the linear regime, the cosmic variance error covariance between two luminosity function bins $i$ and $j$ coming from the matter density field is given by
\begin{equation}
C^{\rm CV}_{ij} = \frac{\phi(L_i)\phi(L_j)}{N_{\rm f}}\int (b_i + f \mu^2)(b_j + f \mu^2) P_{\rm m}(k,z)\,|W({\vec k})|^2\,\frac{\D^3\vec k}{(2\pi)^3},
\label{eq:LSSCov}
\end{equation}
where $N_{\rm f}$ is the number of independent fields, $P_{\rm m}(k,z)$ is the real-space matter power spectrum at redshift $z$, $f$ is the growth rate (which boosts the cosmic variance in narrow-band surveys due to redshift-space distortions), and $W({\vec k})$ is the window function, the Fourier transform of the survey volume, normalized to $W({\vec 0})=1$. We have used this result here assuming a bias of $b_i = \bar b=0.9+0.4z$ (from a semianalytic model, \citealt{2010MNRAS.405.1006O}, although there is evidence that the bias of star-forming galaxies might be higher; see e.g. \citealt{2012MNRAS.426..679G}). This reduces the cosmic variance matrix to
\begin{equation}
C^{\rm CV1}_{ij} = \frac{\phi(L_i)\phi(L_j)}{N_{\rm f}}\int (\bar b+f\mu^2)^2 P_{\rm m}(k,z)\,|W({\vec k})|^2\,\frac{\D^3\vec k}{(2\pi)^3},
\label{eq:LSSCov1}
\end{equation}
(Here all the entries in the covariance are constant.)

The HiZELS error bars do not incorporate a contribution from cosmic variance. However, we can estimate it from Eq.~(\ref{eq:LSSCov}) assuming the geometry of $N_{\rm f}=2$ independent boxes of size $1\times1$ deg each. The depth in the radial direction is given by the width of the narrow-band filter, and is $\Delta z = 0.020$, 0.030, 0.032, and 0.032 at $z=0.40$, 0.84, 1.47, and 2.23 respectively. The faintest bins in HIZELS at $z=2.23$ come from the HAWK-I camera, and the survey volume is smaller in this case: it is a single field, with size $0.125\times 0.125$ deg, and width $\Delta z = 0.046$. For the 4 redshift bins and the luminosity function bins where the full field has been observed, the implied diagonal elements of the covariance are 0.100, 0.045, 0.032, and 0.025. For the HAWK-I data (faintest objects at $z=2.23$), we find a variance of 0.256.

A more subtle issue is that the above procedure assumes that the bias is independent of $L_{{\rm H}\alpha}$. This assumption has been commonly used for the purpose of forecasting H$\alpha$ survey performance and its dependence on survey design. However, in combination with Eq.~(\ref{eq:LSSCov}), it implies that the cosmic variance contributions in each bin are perfectly correlated. That means that a fit using Eq.~(\ref{eq:LSSCov}) will assume that the {\em shape} of the H$\alpha$LF has no cosmic variance: the cosmic variance term will instead allow only the {\em normalization} to float up and down with an uncertainty given by Eq.~(\ref{eq:LSSCov}). Since cosmic variance is the largest contributor to the errors in some luminosity ranges, the procedure above could lead to fit results that are artificially well-constrained, if the bias is in fact dependent on $L_{{\rm H}\alpha}$. There is no reason for $db/d(\log_{10}L_{{\rm H}\alpha})$ to be exactly zero, although for star-forming galaxies it is not obvious which sign to expect. We have thus explored the possibility of averaging the covariance matrix over a range of possible bias models, constrained by some kind of prior. A simple example of such a prior on the bias is that it deviates from the simple fiducial model according to a Markovian process in $\log_{10} L_{{\rm H}\alpha}$,
\begin{equation}
\langle b_i \rangle = \bar b,~~~~
{\rm Cov}(b_i,b_j) = c_1^2\bar b e^{-|\log_{10} L_i-\log_{10} L_j|/c_2},
\end{equation}
which results in a modified cosmic variance term\footnote{There are redshift-space distortion terms in Eq.~(\ref{eq:LSSCov2}) that we have neglected; we do not believe the fidelity of the model warrants a more intricate correction.}
\begin{equation}
C^{\rm CV2}_{ij} = C^{\rm CV1}_{ij} ( 1 + c_1^2 e^{-|\log_{10} L_i-\log_{10} L_j|/c_2} ).
\label{eq:LSSCov2}
\end{equation}
Here $c_1$ is the fractional prior uncertainty in the bias and $c_2$ is its correlation length in $\log_{10}L$. The fiducial parameters taken are $c_1=0.5$ (50\%\ scatter in the bias model) and $c_2=2$ (2 dex correlation length). As always with priors, these parameters are somewhat {\slshape ad hoc}, but despite this drawback we expect that a procedure with a range of bias models is more likely to be able to approximate the real Universe than a fixed-bias case ($C_{ij}=C_{ij}^{\rm CV1}$) or the assumption of no cosmic variance at all.

There are thus 3 possible models for the incorporation of cosmic variance in the narrow-band luminosity function:
\begin{list}{$\bullet$}{}
\item No inclusion of cosmic variance ($C_{ij}=0$).
\item The simple, luminosity-independent bias model ($C_{ij}=C_{ij}^{\rm CV1}$).
\item A random suite of luminosity-dependent bias models ($C_{ij}=C_{ij}^{\rm CV2}$).
\end{list}

The slitless surveys have a very different geometry: they probe tiny areas (e.g. the WFC3 detector covers only 4.8 arcmin$^2$), but they have a very long contribution in the radial direction and usually have many more independent fields ($N_{\rm f}=29$ for WISP). For the $0.3<z<0.9$ and $0.9<z<1.5$ slices, the predicted cosmic variance diagonal covariances for WISP are 0.0020 and 0.0014 respectively. The WISP luminosity function includes the cosmic variance term, although the fitting procedure used here does not include the cosmic variance covariance between luminosity bins. We have not attempted to add these in, as the additional $\sim 4$\%\ standard deviation is negligible.

\section{Poisson error bars}
\label{app:Poisson}

This appendix considers the asymmetry of the Poisson error bar in the context of constructing a likelihood function for the \Ha\ luminosity function for Model 3. The procedure was inspired by applications in cosmic microwave background data analysis, where the anisotropy power spectrum has asymmetric (in that case, $\chi^2$-shaped) error bars \citep{2003ApJS..148..195V}.
A common example is in power spectrum estimation, where the overall fit can be biased downward if symmetric error bars are assumed because the lower data points have smaller error bars and pull the fit. For this reason, parameterized forms of the asymmetry are common in reporting likelihood functions in the cosmic microwave background community (see e.g. \citealt{1998PhRvD..57.2117B, 2000ApJ...533...19B,2003ApJS..148..195V}). A similar phenomenon can occur in fitting a luminosity function: the Poisson error bar on a data point that fluctuates downward is smaller than on a point that fluctuates upward, so fits to the raw luminosity function that treat this error as symmetric will be biased toward lower $\phi(L,z)$. As an extreme example, the likelihood function will even allow a finite likelihood for $\phi(L,z)<0$, which is clearly unphysical. On the other hand, treating the error on $\log_{10}\phi(L,z)$ as symmetric will bias $\phi(L,z)$ upward, since data points that fluctuate upward will have smaller error bars in log-space.

If the H$\alpha$ luminosity function measurements contained only Poisson errors, then the log-likelihood for a point with $N$ objects, a survey volume $\Delta V$, and a bin width $\Delta L$ is
\begin{equation}
\ln {\cal L} = -\ln(N!) - \lambda + N\ln\lambda,
\end{equation}
where $\lambda = \phi\,\Delta L\,\Delta V$ is the expected number of objects. The maximum likelihood point is at $\lambda=N$, and so the log-likelihood relative to the maximum is
\begin{equation}
\ln {\cal L} - \ln{\cal L}_{\rm max} = N\left( 1 - \frac\lambda N+ \ln\frac{\lambda}{N} \right).
\end{equation}
The estimate of the luminosity function is $\hat\phi = N/(\Delta L\,\Delta V)$, and the estimate of its uncertainty is $\sigma_{\ln\phi} = 1/\sqrt{N}$, so this can be re-written as
\begin{equation}
\ln {\cal L} - \ln{\cal L}_{\rm max} = \frac{1}{\sigma_{\ln\phi}^2}\left( 1 - \frac\phi{\hat\phi}+ \ln\frac{\phi}{\hat\phi} \right) = - \frac{x^2}{2\sigma_{\ln\phi}^2},
\end{equation}
where we have defined the re-scaled parameter $x$ as follow:
\begin{equation}
x = \pm \sqrt{2\left( \frac\phi{\hat\phi}-1- \ln\frac{\phi}{\hat\phi} \right)},
\label{eq:xdef}
\end{equation}
with the $+$ sign used if $\phi>\hat\phi$ and the $-$ sign if $\phi<\hat\phi$. We note that the argument of the square root is always positive (or 0 if $\phi=\hat\phi$), and that $x$ is actually an analytic function of $y=\phi/\hat\phi-1$,
\begin{equation}
x = \pm\sqrt{2[y-\ln(1+y)]} = y - \frac13y^2 + \frac7{36}y^3 - ...
\label{eq:xy-P}
\end{equation}
The real error bars need not have the same asymmetry as the Poisson distribution in the cases where they are dominated by other terms (e.g. cosmic variance). We therefore test for the sensitivity of the results to the assumed fitting scheme.

The covariance matrix ${\tens C}$ is re-written in terms of $x$, and the log-likelihood surface is taken to be quadratic,
\begin{equation}
\chi^2 = -2\ln{\cal L} + 2\ln{\cal L}_{\rm max} = \sum_{ij} [{\tens C}^{-1}]_{ij} x_i x_j.
\end{equation}
This approach has the advantage that with one switch in the fitting code, the error asymmetry may be treated in 4 ways:
\begin{list}{$\bullet$}{}
\item {\em Poisson}: This uses Poisson-shaped errors (Eq.~\ref{eq:xy-P}).
\item {\em Symmetric-linear}: This uses symmetric errors in $\phi$, by setting $x=y$.
\item {\em Symmetric-log}: This uses symmetric errors in $\log_{10}\phi$ or $\ln\phi$, by setting $x=\ln(1+y)$.
\item {\em Symmetric-native}: This uses errors symmetric in either $\phi$ or $\ln\phi$, depending on which was reported by the analysis team.
\end{list}
The Poisson shape for the error bars is probably the most realistic in the bins with small numbers of galaxies, but due to the contribution of other errors it is not exact. Therefore we consider other shapes as well (see Appendix~\ref{app:M3}).

\section{Variations and robustness of Model 3}
\label{app:M3}

\begin{figure}
\includegraphics[width=88mm]{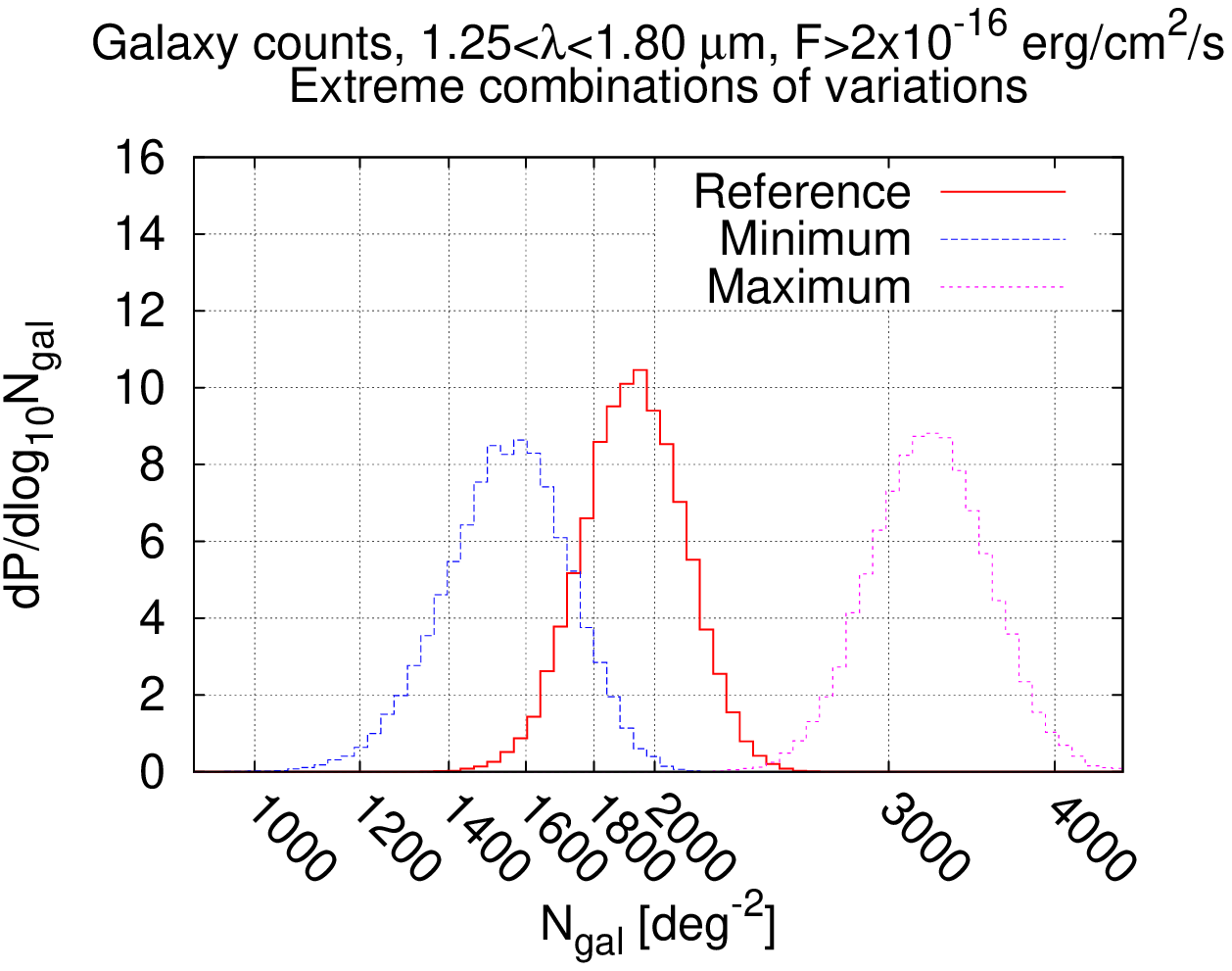}
\caption{\label{fig:e2num} Posterior probability distribution for the number of galaxies at $F>2\times 10^{-16}$ erg cm$^{-2}$ s$^{-1}$ and in the wavelength range 1.25--1.80 $\mu$m (redshift 0.90--1.74), for reference Model 3 and its extreme combinations of modifications considered (see text).}
\end{figure}

\begin{table*}
\caption{\label{tab:pars} Fit parameters for the various models considered. Central values are for the maximum likelihood model, and error ranges shown are 95 percent enclosed posterior intervals (i.e. 2$\sigma$). Of the remainder, 2.5\%\ of the posterior is at lower values and 2.5\%\ at higher values (except for values marked with a $\star$, which indicate a one-sided error bar; these are chosen where the extreme legal value of a parameter, e.g. $\beta=0$ or $\gamma=1$, is allowed). The final column ($N_2$) is the number of galaxies per square degree with an H$\alpha$ line in the range 1.25--1.80 $\mu$m with a flux exceeding $2\times 10^{-16}$ erg cm$^{-2}$ s$^{-1}$. Units are Mpc$^{-3}$ ($\phi_\star$) and erg s$^{-1}$ ($L_\star$).} 
\small{
\begin{tabular}{lrrrrrrrr}
\hline\hline\multicolumn{9}{c}{Reference parameters} \\ 
& $\alpha$ & $\Delta$ & $\log_{10}\phi_{\star,0}$ & $\log_{10}L_{\star,2.0}$ & $\log_{10}L_{\star,0.5}$ & $\beta$ & $\chi^2$/dof & $N_2$ \\[1ex]
{\tt REF} & $-1.587^{+0.132}_{-0.119}$ & $ 2.288^{+0.410}_{-0.379}$ & $-2.920^{+0.183}_{-0.175}$ & $42.557^{+0.109}_{-0.119}$ & $41.733^{+0.150}_{-0.142}$ & $ 1.615^{+0.947}_{-1.196}$ & $ 64.06$/76 & $1950^{+330}_{-330}$ \\[1ex]
\hline\multicolumn{9}{c}{Alternate functional forms} \\ 
& $\alpha$ & $\gamma$ & $\log_{10}\phi_{\star,0}$ & $\log_{10}L_{\star,\infty}$ & $\log_{10}L_{\star,0.5}$ & $\beta$ & $\chi^2$/dof & $N_2$ \\[1ex]
{\tt hybrid} & $-1.555^{+0.158}_{-0.108}$ & $\star 1.000_{-0.402}$ & $-2.851^{+0.206}_{-0.154}$ & $42.871^{+1.125}_{-0.305}$ & $41.689^{+0.136}_{-0.166}$ & $ 1.699^{+1.071}_{-1.062}$ & $ 66.40$/76 & $2022^{+329}_{-314}$ \\[1ex]
& $\alpha$ & $\log_{10}\phi_{\star,1}$ & \!$(d/da)\log_{10}\phi_{\star,0}$ & $\log_{10}L_{\star,\infty}$ & $\log_{10}L_{\star,0.5}$ & $\beta$ & $\chi^2$/dof & $N_2$ \\[1ex]
{\tt schechter} & $-1.526^{+0.103}_{-0.184}$ & $-2.752^{+0.124}_{-0.303}$ & $-0.018^{+0.491}_{-1.297}$ & $42.857^{+3.139}_{-0.277}$ & $41.647^{+0.406}_{-0.138}$ & $ 1.655^{+0.957}_{-1.425}$ & $ 83.96$/76 & $2100^{+318}_{-341}$ \\[1ex]
\hline\multicolumn{9}{c}{Extreme cases} \\ 
& $\alpha$ & $\Delta$ & $\log_{10}\phi_{\star,0}$ & $\log_{10}L_{\star,2.0}$ & $\log_{10}L_{\star,0.5}$ & $\beta$ & $\chi^2$/dof & $N_2$ \\[1ex]
{\tt MIN} & $-1.656^{+0.129}_{-0.106}$ & $ 2.916^{+0.718}_{-0.598}$ & $-3.039^{+0.180}_{-0.156}$ & $42.583^{+0.092}_{-0.124}$ & $41.772^{+0.127}_{-0.155}$ & $ 1.698^{+1.333}_{-1.180}$ & $ 28.31$/76 & $1596^{+283}_{-359}$ \\[1ex]
{\tt MAX} & $-1.385^{+0.255}_{-0.229}$ & $ 1.598^{+0.329}_{-0.326}$ & $-2.690^{+0.322}_{-0.373}$ & $42.539^{+0.245}_{-0.343}$ & $41.781^{+0.271}_{-0.291}$ & $\star 0.010^{+1.761}$ & $ 25.91$/76 & $3169^{+770}_{-533}$ \\[1ex]
\hline\hline
\end{tabular}
}
\end{table*}
In order to assess the robustness of Model 3, we re-ran the fits modifying some of the key aspects of the data handling.
The reference model 
is based on (i) use of all data sets; (ii) the broken power law model for the luminosity function; (iii) the CV2 cosmic variance prescription; (iv) the Poisson error bar asymmetry model; (v) integration over luminosity and redshift bins using $N_{\rm G}=3$; (vi) HiZELS aperture corrections assuming a 0.3 arc sec half-light radius for all sources; and (vii) the \nii/\Ha\ ratio assumed in the input publications.
We vary the reference assumptions (functional function, CV, $N_{\rm G}$, error bars) and we also considered extreme combinations of modifications to come up with bounding optimistic ({\tt MAX}) or conservative ({\tt MIN}) estimates of the H$\alpha$LF. Relative to the reference fit, the {\tt MAX} fit used the combination of fits to the bin centre; error bars symmetric in $\log\phi$; and WISP+NICMOS data only. The {\tt MIN} fit used the combination of error bars symmetric in $\phi$; and HiZELS+WISP data only.
The main types of variations considered, and results of the fit are listed in Table~\ref{tab:pars}.
We consider the predictions of the models for the number of galaxies $N_2$ above $2\times 10^{-16}$ erg cm$^{-2}$ s$^{-1}$ and in the H$\alpha$ 
redshift range (0.9<z<1.74).

The fits with more simplistic treatment of the finite bin width 
(using $N_{\rm G}=1$ and the luminosity function at $z=(z_{\rm min}+z_{\rm max})/2$ and $\log_{10} L=(\log_{10}L_{\rm min} + \log_{10}L_{\rm max})/2$)
lead to higher predicted counts. This is the result of Eddington-like biases: for a steeply falling luminosity function\footnote{Technically, one with large second derivative.}, a bin of width $\Delta\log_{10}L \times \Delta z$ contains more galaxies than would be predicted based on the luminosity function at the bin centre. The reference fit corrects this effect by incorporating it in the model. 
The $N_{\rm G}=5$ case was run as a convergence test, and shows $\ll 1\sigma$ changes. 
The differences between the cases indicate the significance of different ways of treating finite bin size. The uncertainties are largest for the NICMOS data since large bins in both $\log L$ and $z$ were used in the NICMOS studies \citep{1999ApJ...519L..47Y,2009ApJ...696..785S}. The effect of this treatment is smallest for HiZELS since there is no averaging over redshifts and the $\log L$ bins are narrow.

The choice of cosmic variance treatment (CV1 versus CV2) matters little
($\ll 1\sigma$) in the integrated counts in the {\slshape Euclid} range $F_{{\rm H}\alpha}>2\times 10^{-16}$ erg cm$^{-2}$ s$^{-1}$ from switching between these two models, although the faint-end slope changes by $1\sigma$.

A bigger difference arises 
when the cosmic variance is artificially turned off; this causes the predicted number of galaxies to go up by $2\sigma$. This behaviour is driven by the three lowest-luminosity HiZELS points at $z=0.84$, which have small formal error bars (0.03 or 0.04 dex) and are actually above the WISP counts. 

The treatment of error bar asymmetries pulls the fits in the expected direction: treating the error bars as symmetric in $\phi$ leads to a lower result by almost $2\sigma$, and treating them as symmetric in $\log_{10}\phi$ leads to a higher result by almost $2\sigma$, relative to the Poisson-shaped error bar. The Poisson shape (reference) is the best-motivated form, since we know that a major contribution to the luminosity function error has this shape, but many past fits have been done with one of the two other shapes, and we do not have a clear understanding of the asymmetry of the systematic errors.

We performed fits excluding each of the 3 major input samples 
Since the narrow-band HiZELS H$\alpha$ luminosity function is the lowest in the {\slshape Euclid} range, and the NICMOS results are the highest, exclusion of HiZELS moves the predicted number of galaxies up, whereas exclusion of NICMOS moves it down. The difference between the highest and lowest result in this sample jack-knife is 0.161 dex.
This suggests that systematic errors are contributing to the differences of these curves and that caution should be exercised in interpreting joint fits.

The alternative fitting functions, especially the Schechter function, lead to slightly greater number densities than the reference (broken power law). This is because they incorporate an exponential cutoff, and hence the existence of a few very bright galaxies ($>5L_\star$, particularly in the NICMOS data) pulls the characteristic luminosity to larger values and increases the number of objects in the intermediate range ($\sim 2L_\star$). However, this same feature of the Schechter law means that it is a poor fit to the NICMOS observations, and it is disfavoured relative to the broken power law model by $\Delta\chi^2=20$, and in any case the effect in our fiducial range ($F_{{\rm H}\alpha}>2\times 10^{-16}$ erg cm$^{-2}$ s$^{-1}$, $0.90<z<1.74$) is only $1\sigma$.

The reference aperture correction for HiZELS assumes a half-light radius of 0.3 arc sec, which is consistent with objects near the flux limit of {\slshape WFIRST-AFTA} (see \citealt{2013ApJ...779...34C}, Fig. 11). We have tried two variations on this: an extreme case of turning the aperture correction off, 
and a case of implementing a variable galaxy size in accordance with the fit provided in \S4.2 of \citet{2013ApJ...779...34C}.
\footnote{For this fit, the \Ha\ luminosities were re-scaled, and the differential luminosity function was appropriately transformed using the Jacobian of the uncorrected-to-corrected flux transformation.} The changes in the number of objects in the range $0.90<z<1.74$ and at $F_{{\rm H}\alpha}>2\times 10^{-16}$ erg cm$^{-2}$ s$^{-1}$ are $-10$\% and $-2$\% for the no aperture correction and \citet{2013ApJ...779...34C} correction 
cases, respectively.

The last modelling assumption that was varied was the assumed \nii/\Ha\ ratio, which enters because at low resolution \nii\ and \Ha\ are blended; thus \Ha+\nii\ is measured, and \Ha\ is inferred under some assumed prescription for the line ratio. The reference model is based on the \Ha\ luminosity function directly from the published papers: this means that the assumed \nii/\Ha\ is that in the published papers (0.41 for NICMOS and WISP; in HiZELS a variable ratio was used but the reported median is 0.33). Here \nii\ includes both doublet members, 6548 \AA\ and 6583 \AA; 75.4\%\ of the flux in the stronger 6583 \AA\ line \citep{2000MNRAS.312..813S}. This ratio is common at low redshifts, however a range of values is observed, and in high-redshift galaxies the \nii/\Ha\ ratio is often observed to be smaller. We have therefore investigated what happens under alternate assumptions regarding the \nii/\Ha\ ratio. First, the luminosities were converted back to $L_{{\rm H}\alpha + [{\rm NII}]}$ using the stated median ratios in each input paper. Then the \Ha+\nii\ luminosity function was written as
\begin{equation}
\phi_{{\rm H}\alpha + [{\rm NII}]}(L_{{\rm H}\alpha + [{\rm NII}]}) = \int \phi_{{\rm H}\alpha}(L_{{\rm H}\alpha})
p(x|L_{{\rm H}\alpha}) \left.\frac{\partial L_{{\rm H}\alpha}}{\partial x}\right|_{L_{{\rm H}\alpha + [{\rm NII}]}} \,dx,
\label{eq:intquad}
\end{equation}
where $x = \log_{10}(L_{6583}/L_{{\rm H}\alpha})$ is the relative line strength in dex and $L_{{\rm H}\alpha} = L_{{\rm H}\alpha + [{\rm NII}]} / (1 + 10^x/0.754)$. We built two alternative models for the \nii/\Ha\ ratio based on the $\langle z\rangle = 2.3$ BPT diagram of star-forming galaxies \citep{2014ApJ...795..165S}. One model ({\tt altNII1}) uses the median \nii/\Ha\ ratio from the \citet{2014ApJ...795..165S} sample, $x=-0.90$ dex (see Figure 5). The other ({\tt altNII2}) assumes a lognormal distribution; since the 84th percentile ($+1\sigma$) of the \nii/\Ha\ ratio corresponds to $x=-0.57$ dex, we choose a median at $-0.90$ dex and a scatter of $\sigma_x = 0.33$ dex. 

In the {\tt altNII1} model, the number of objects in the range $0.90<z<1.74$ and at $F_{{\rm H}\alpha}>2\times 10^{-16}$ erg cm$^{-2}$ s$^{-1}$ increases 
by 45\%; the weaker assumed \nii\ results in larger inferred \Ha\ luminosities, and this effect is amplified by the steep luminosity function. On the other hand, for the {\tt altNII2} model, which includes scatter as well, we find a source density 
only 29\%\ above the reference model; the reduction occurs because the scatter in \nii\ results in an Eddington-like bias that is corrected by Eq.~(\ref{eq:intquad}). While an improvement over the reference model in some ways, the 29\%\ increase in the {\tt altNII2} model may be an overestimate, since (i) it applies a correction based on the $\langle z\rangle = 2.3$ BPT diagram even at lower redshifts, and (ii) the correction procedure is not technically correct for HiZELS, which has a variable assumed \nii\ fraction and which may include only part of the \Ha+\nii\ complex in its band.\footnote{The correction in Eq.~(\ref{eq:intquad}) is an overestimate in cases where \Ha\ falls in the narrow bandpass and one or both of the \nii\ lines do not. It is an underestimate if \nii\ 6583 \AA\ falls in the narrow band and \Ha\ does not, but since \Ha\ is almost always stronger this is not as much of an issue at the top of the luminosity function.} There may also be differences (whose impact has undetermined sign) between the rest-frame ultraviolet selection in \citet{2014ApJ...795..165S} and \Ha\ selection. Based on these considerations, 
we are not using it to replace the reference model.

Finally, it is seen that the central values of the {\tt MIN} and {\tt MAX} fits for the number of objects in the range $0.90<z<1.74$ and at $F_{{\rm H}\alpha}>2\times 10^{-16}$ erg cm$^{-2}$ s$^{-1}$ differ by a factor of 2
(see Table~\ref{tab:pars}). 

\end{appendix}

\end{document}